\documentclass{jfm}
\usepackage{bm} 
\usepackage{amssymb}
\usepackage{graphicx}
\usepackage{color}
\usepackage{amsmath}
\usepackage{amsfonts}
\usepackage{bm}
\usepackage{latexsym}
\usepackage{epsfig}
\newcommand{\la}{\langle}
\newcommand{\ra}{\rangle}

\newcommand{\cP}{{\cal P}}
\newcommand{\cW}{{\cal W}}
\newcommand{\ee}{\end{equation}}

\title[Turbulent pair dispersion of inertial particles]{Turbulent pair
  dispersion of inertial particles}

\author[J.\ Bec, L.\ Biferale, A.\ S.\ Lanotte, A.\ Scagliarini, and
  F.\ Toschi] {J.\ns B\ls E\ls C$^1$, L.\ns B\ls I\ls F\ls E\ls R\ls
  A\ls L\ls E$^{2}$, A.\ns S.\ns L\ls A\ls N\ls O\ls T\ls T\ls
  E$^{3}$, A.\ns S\ls C\ls A\ls G\ls L\ls I\ls A\ls R\ls I\ls N\ls
  I$^{2}$ and F.\ns T\ls O\ls S\ls C\ls H\ls I$^{4}$}

\affiliation{$^1$ Universit\'{e} de Nice-Sophia Antipolis, CNRS,
  Observatoire de la C\^{o}te d'Azur,\\ Laboratoire Cassiop\'{e}e,
  Bd.\ de l'Observatoire, 06300 Nice, France \\[\affilskip] $^2$
  Department of Physics and INFN, University of Rome Tor Vergata, \\
  Via della Ricerca Scientifica 1, 00133 Roma, Italy \\ [\affilskip]
  $^3$ ISAC-CNR, Istituto di Scienze dell'Atmosfera e del Clima, Via
  Fosso del Cavaliere 100, 00133 Roma and INFN, Sezione di Lecce,
  73100 Lecce Italy \\ [\affilskip] $^4$ Department of Physics and
  Department of Mathematics and Computer Science, Eindhoven University
  of Technology, 5600 MB Eindhoven, The Netherlands and Istituto per
  le Applicazioni del Calcolo CNR, Viale del Policlinico 137, 00161
  Roma, Italy}

\begin{document}

\maketitle

\begin{abstract}
  The relative dispersion of pairs of inertial particles in
  incompressible, homogeneous, and isotropic turbulence is studied by
  means of direct numerical simulations at two values of the
  Taylor-scale Reynolds number $Re_{\lambda} \sim 200$ and
  $Re_{\lambda} \sim 400$, corresponding to resolutions of $512^3$ and
  $2048^3$ grid points, respectively. The evolution of both heavy and
  light particle pairs is analysed at varying the particle Stokes
  number and the fluid-to-particle density ratio. For particles much
  heavier than the fluid, the range of available Stokes numbers is $St
  \in [0.1\!:\!70]$, while for light particles the Stokes numbers span
  the range $St \in [0.1\!:\!3]$ and the density ratio is varied up to
  the limit of vanishing particle density.\\ For heavy particles, it
  is found that turbulent dispersion is schematically governed by two
  temporal regimes. The first is dominated by the presence, at large
  Stokes numbers, of small-scale caustics in the particle velocity
  statistics, and it lasts until heavy particle velocities have
  relaxed towards the underlying flow velocities. At such large
  scales, a second regime starts where heavy particles separate as
  tracers particles would do. As a consequence, at increasing inertia,
  a larger transient stage is observed, and the Richardson diffusion
  of simple tracers is recovered only at large times and large
  scales. These features also arise from a statistical closure of the
  equation of motion for heavy particle separation that is proposed,
  and which is supported by the numerical results.\\ In the case of
  light particles with high density ratios, strong small-scale
  clustering leads to a considerable fraction of pairs that do not
  separate at all, although the mean separation increases with
  time. This effect strongly alters the shape of the probability
  density function of light particle separations.
\end{abstract}

\section{Introduction}
\label{sec:intro}

Suspensions of dust, droplets, bubbles, and other finite-size
particles advected by incompressible turbulent flows are commonly
encountered in many natural phenomena (see, e.g.,
\cite{Csanady,Eaton,falko_nature,Abraham,Shaw,tbreview}).
Understanding their statistical properties is thus of primary
importance. From a theoretical point of view, the problem is more
complicated than in the case of fluid tracers, i.e.\ point-like
particles with the same density as the carrier fluid. Indeed, when the
suspended particles have a finite size and a density ratio different
from that of the fluid, they have inertia and do not follow exactly
the flow. As a consequence, correlations between particle positions
and structures of the underlying flow appear. It is for instance well
known that heavy particles are expelled from vortical structures,
while light particles tend to concentrate in their cores. This results
in the formation of strong inhomogeneities in the particle spatial
distribution, an effect often refered to as {\it preferential
  concentration} (see \cite{Douady,SE91,Eaton}).  This phenomenon has
gathered much attention, as it is revealed by the amount of recently
published theoretical work (\cite{Falkovich-clustering,simo,
  Falkovich-Pumir}), and numerical studies
(\cite{Collins1,Collins2,prl_nostro,vassilicos}).  Progresses in the
statistical characterization of particle aggregates have been achieved
by studying particles evolving in stochastic flows by
\cite{stuart,mehlig-wilkinson,bccm05,olla} and in two-dimensional
turbulent flows by \cite{Boffetta}. Also, single trajectory statistics
have been addressed both numerically and experimentally for small
heavy particles (see, e.g.,
\cite{bec.jfm,bec.jot,warhaft,gerashenko,zaichik,warhaft2,volk}), and
for large particles (\cite{bourgoin,bodi}). The reader is refered to
\cite{tbreview} for a review.

In this paper we are concerned with particle pair dispersion, that is
with the statistics, as a function of time, of the separation distance
$\bm R(t) = \bm X_1(t)-\bm X_2(t)$ between two inertial particles, labelled
by the subscripts 1 and 2 (see \cite{cencini,fouxon,derevich} for
recent studies on that problem). In homogeneous turbulence, it is
sufficient to consider the statistics of the instantaneous separation
of the positions of the two particles. These are organised in
different families according to the values of their Stokes number
$St$, and of their density mismatch with the fluid, $\beta$.

For our purposes, the motion of particle pairs, with given
$(St,\beta)$ values and with initial separations inside a given
spherical shell, $ R = |\bm X_1(t_0)-\bm X_2(t_0)| \in [R_0,R_0+dR_0]$
is followed until particle separation reaches the large scale of the
flow. With respect to the case of simple tracers, the time evolution
of the inertial particle pair separation $R(t)$ becomes a function not
only of the initial distance $R_0$, and of the Reynolds number of the
flow, but also of the inertia parameters $(St,\beta)$.

A key question that naturally arises is how to choose the initial
spatial and velocity distributions of inertial pairs. Indeed, it is
known that heavy (resp.\ light) particles tend to concentrate
preferentially in hyperbolic (resp.\ elliptic) regions of the
advecting flow, with spatial correlation effects that may extend up to
the inertial range of scales, as shown in \cite{prl_nostro}.
Moreover, when inertia is high enough, the particle pair velocity
difference, $\delta_R V = |\bm V_1(\bm X_1(t),t)-\bm V_2(\bm
X_2(t),t)|$, may not go smoothly to zero when the particle separations
decreases, a phenomenon connected to the formation of \emph{caustics},
see \cite{caus2,caus1}. In our numerical simulations, particles of
different inertia are injected into the flow and let evolve until they
reach a stationary statistics for both spatial and velocity
distributions. Only after this transient time, pairs of particles with
fixed intial separation are selected and then followed in the spatial
domain to study relative dispersion.

By reason of the previous considerations, the main issue is to
understand the role played by the spatial inhomogeneities of the
inertial particle concentration field and by the presence of caustics
on the pair separations, at changing the degree of inertia. We remark
that these two effects can be treated as independent only in the limit
of very small and very large inertia. In the former case, particles
tend to behave like tracers and move with the underlying fluid
velocity: preferential concentration may affect only their
separation. In the opposite limit, particles distribute almost
homogeneously in the flow: however, due to their ballistic motion,
they can reach nearby positions with very different velocities
(\cite{falko_nature}). In any other case of intermediate inertia, both
these effects are present and may play a role in the statistics of
inertial pair separation.

It is worth anticipating the two main results of this
study:\\ (\textit{i}) The separation between heavy particles can be
described in terms of two time regimes: a first regime is dominated by
inertia effects, and considerable deviations from the tracers case
arise in the inertial relative dispersion; in the second one, the
tracers behaviour is recovered since inertia is weak and appears only
in subdominant corrections that vanish as $1/t$. The crossover between
these two regimes defines a new characteristic spatial and temporal
scale, connected to both the \emph{size of caustics} and the Stokes
number, which influences the particle separation for not too long
time-lags and not too large scale.\\ (\textit{ii}) The strong
clustering properties that are typical of light particles may lead to
the fact that many pairs do not separate at all: their statistical
weight is clear in the separation probability density function (PDF),
which develops a well defined power-law left tail.

It would clearly be also interesting to investigate the dependence
upon the Reynolds number of the inertial particle pair
separation. Small-scale clustering seems to be poorly dependent on the
degree of turbulence of the carrier flow (\cite{Collins1,prl_nostro}),
while much less is known about the Reynolds number dependence of the
caustics statistics. Our numerical data do not allow to explore this
question in detail, so that we will restrict ourselves to show data
associated to the two Reynolds numbers in all cases when differences
are not significative.

In the case of fluid tracers, the standard observables are the time
evolutions of the mean square separation and of the separation
probability density function, for which well established predictions
exist since the pioneering work of \cite{rich.rev}. We contrast these
observables obtained for tracers with the results for heavy and light
inertial particles.

The paper is organised as follows. In \S\ref{sec:0}, we briefly recall
the basic equations of motion and describe the numerical
simulations. In \S\ref{sec:1}, we analyse the stationary distribution
of heavy particle velocity differences, conditioned on the particle
initial separation, highlighting both the presence of small-scale
caustics and the effects of particle inertia at those scales
corresponding to the inertial range of turbulence. In \S\ref{sec:2} we
study the behaviour of the mean separation distance of heavy pairs, at
changing the Stokes number $St$; we also analyse the influence of the
caustics in the initial statistics on the subsequent pair separation
evolution. A {\it mean-field} model, which is able to capture the main
numerical findings, is proposed in the same section.  The time
evolution of the separation probability density functions is discussed
in \S\ref{sec:pdf} and we present the data for light particles in
\S\ref{sec:light}. In \S\ref{sec:conc} we summarise the main findings.

\section{Equation of motion and numerical details}
\label{sec:0}
We present results from direct numerical simulations of turbulent
flows seeded with inertial particles. The flow phase is described by
the Navier-Stokes equations for the velocity field ${\bm u}(\bm x,t)$
\begin{equation}
  \partial_t\bm u + \bm u \cdot \nabla \bm u = -\nabla p +
  \nu\nabla^2\bm u +\bm f,\quad \nabla\cdot\bm u = 0\,.
  \label{eq:ns}
\end{equation}
The statistically homogeneous and isotropic external forcing $\bm f$
injects energy in the first low wave number shells, by keeping
constant their spectral content (see \cite{She}).
\begin{table*}
  \centering
  \begin{tabular}{cccccccccc} \hline\\[-13pt]
      & $N$ & $Re_{\lambda}$ & $\eta$ & $\delta x$ & $\varepsilon$ &
    $\nu$ & $\tau_{\eta}$ & $t_{\mathrm{dump}}$ & $\delta t$
    \\[-5pt]\hline Run I& 512 & 185 & 0.01 & 0.012 & 0.9 & 0.002 &
    0.047 & 0.004 & 0.0004 \\ Run II& 2048 & 400 & 0.0026 & 0.003 &
    0.88 & 0.00035 & 0.02 & 0.00115 & 0.000115\\[-5pt]\hline
  \end{tabular}
  \caption{\label{table} Eulerian parametres for the two runs analysed
    here: Run I and Run II in the text. $N$ is the number of grid
    points in each spatial direction; $Re_{\lambda}$ is the
    Taylor-scale Reynolds number; $\eta$ is the Kolmogorov dissipative
    scale; $\delta x=\mathcal{L}/N$ is the grid spacing, with
    $\mathcal{L}=2\pi$ denoting the physical size of the numerical
    domain; $\tau_\eta =\sqrt{\nu/\varepsilon}$ is the Kolmogorov
    dissipative time scale; $\varepsilon$ is the kinetic energy
    dissipation; $\nu$ is the kinematic viscosity;
    $\tau_{\mathrm{dump}}$ is the time interval between two successive
    dumps along particle trajectories; $\delta t$ is the time step.}
\end{table*}
The kinematic viscosity $\nu$ is chosen such that the Kolmogorov
length scale $\eta\approx \delta x$, where $\delta x$ is the grid
spacing: this choice ensures a good resolution of the small-scale
velocity dynamics. The numerical domain is cubic and $2\pi$-periodic
in the three directions of space. We use a fully dealiased
pseudospectral algorithm with 2$^{\mathrm{nd}}$ order Adam-Bashforth
time-stepping (for details see \cite{bec.jfm,bec.jot}). We performed
two series of DNS: Run I with numerical resolution of $512^3$ grid
points, and the Reynolds number at the Taylor scale $Re_\lambda
\approx 200$; Run II with $2048^3$ resolution and $Re_\lambda \approx
400$. Details of the runs can be found in Table~\ref{table}.

The particle phase is constituted by millions of heavy and light
particles\,---\,the latter only for Run I\,---\,with different
intrinsic characteristics. Particles are assumed to be with size much
smaller than the Kolmogorov scale of the flow, $\eta$, and with a
negligible Reynolds number relative to the particle size. In this
limit, the equations ruling their dynamics take the particularly
simple form:
\begin{equation}
\dot {\bm X} \, =\, \bm V \; , \qquad \dot{\bm V} \,=\,
-\frac{1}{\tau_s}\left[\bm V-{\bm u}(\bm X,t)\right] + \beta\, D_t {\bm
  u}(\bm X,t) \;,
\label{eq:1}
\end{equation}
where the dots denote time derivatives. The particle position and
velocity are $(\bm X(t),\bm V(t))$, respectively; ${\bm u}({\bm
  X}(t),t)$ is the Eulerian fluid velocity evaluated at the particle
position, and $D_t {\bm u}$ is the so-called added mass term, which
measures the fluid acceleration along particle trajectory.  The
adimensional constant $\beta= 3\rho_f/(\rho_f + 2 \rho_p)$ accounts
for the added mass effect through the density contrast between
particles $\rho_p$ and fluid $\rho_f$. The particle response time,
appearing in the Stokes drag, is $\tau_s =2 \rho_p a^2 /(9 \rho_f
\nu)$, where $a$ is the particle radius. Particle inertia is
quantified by the \emph{Stokes number} that is defined as
$St=\tau_s/\tau_\eta$, where $\tau_\eta=(\nu/\varepsilon)^{1/2}$ is
the flow Kolmogorov timescale and $\varepsilon$ the average rate of
energy injection. Equation (\ref{eq:1}) has been derived in
\cite{maxey2} under the assumption of very dilute suspensions, where
particle-particle interactions (collisions) and hydrodynamic coupling
to the flow can be neglected.

For Run I, we show results for the following set of $(St,\beta)$
families: (\textit{i}) very heavy particles [$\beta=0$]:
$St=0.0,0.6,1.0,3.3$; (\textit{ii}) light particles [$\beta=2,3$]:
$St=0.3,1.2,4.1$. For each family the typical number of particle pairs
that are followed is around $5 \times 10^4$. For Run II, we show
results only for heavy particles but with a larger range of variation
in the Stokes number: $St=0.0,0.6,1.0,3.0,10,30,70$. Typical number of
particle pairs for each family is $\sim 10^4$. Once injected particles
have relaxed to their steady-state statistics, pairs have been
selected with the following initial separations: $R_0 \le \eta$ and
$R_0 \in [4\!:\!6] \eta$ for both Run I and Run II, and $R_0 \in
[9\!:\!11] \eta$ for Run II only.

Beside the time evolution of particle pairs, we also have
instantaneous snapshots of the two phases (fluid and dispersed), with
a much higher particle statistics: around $10^6$ per family for Run I,
and $10^8$ per family for Run II. These are used to measure the
stationary\,---\,i.e.\ not along the trajectories\,---\,distribution
of particle velocity increments discussed in next section.

\section{Stationary distributions: velocity increments conditioned
 on particles separation}
\label{sec:1}
Turbulent pair dispersion for tracers is classically based on the
application of similarity theory for Eulerian velocity statistics:
depending on the value of space and time scales, velocity increment
statistics differently affect the way tracers separate. This results
in different regimes for relative dispersion, see e.g. \cite{sawford}.  

In the case of inertial particles, the same reasoning holds, so that
to analyse the way inertial pairs separate in time, the stationary
statistics of particles velocity differences has to be investigated
first. A stationary distribution for the typical velocity differences
between two inertial particles is obtained by imposing periodic
boundary conditions inside the physical volume and then measuring
velocities on such a thermalised configuration.

We are interested in the scaling behaviour of velocity increments at
varying the degree of inertia and the distance between the particles
(in the dissipative or inertial range of the turbulent fluid flow).
To fix the notation, we denote by $U_0$ the typical large-scale
velocity of the fluid tracers and by $L$ the integral scale of the
flow. Moreover we define
\begin{equation}
\delta_R V_{St} \,=\, |\bm V_1(\bm X_1(t))-\bm V_2(\bm X_2(t))|,
\label{def:0}
\end{equation}
as the velocity difference at scale $R$, conditioned on the presence
of a pair of particles with Stokes number $St$, separated with a
distance $R = |\bm X_1(t_0)-\bm X_2(t_0)|$. Since we are here interested
in the case of heavy particles only, the Stokes number is sufficient
to identify a given particle family. For convenience, we introduce a
specific notation for the tracer stationary velocity statistics:
$\delta_R u \,=\, \delta_R V_{(St=0)}$, which is exactly equal to the
Eulerian velocity increment at scale $R$.

\begin{figure*}
\begin{center}
\includegraphics[width=.7\textwidth]{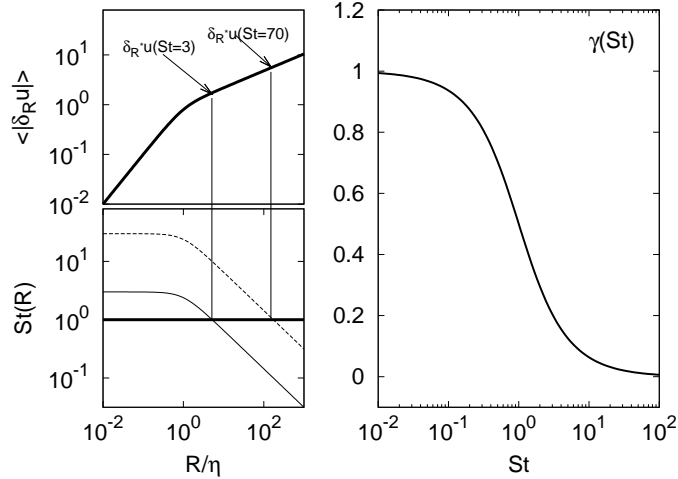}
\caption{Left panels. Bottom figure: behaviour of the scale-dependent
  Stokes number, $St(R)$ as a function of the scale $R$ normalised
  with the Kolmogorov scale $\eta$, for two Stokes numbers $St=3,70$
  (bottom and top, respectively). The horizontal thick line is for
  $St(R)=1$. Top figure: the scaling behaviour for the fluid tracer
  velocity increments versus the scale as given by (\ref{eq:deltau}).
  Notice that the scales $R^*$ where $St(R^*) =1$ fall in the inertial
  range of the Eulerian fluid velocity. Right panel: the function
  $\gamma(St)$ defining the small scales power law behaviour of
  caustics statistics at changing inertia. Notice that for small
  values of the Stokes number $St$, $\gamma \rightarrow 1$,
  i.e.\ particle velocity is differentiable; at high inertia,
  $\gamma \rightarrow 0$ indicating the existence of discontinuities
  in the particle velocity increment statistics.}
\label{fig:stokes}
\end{center}
\vspace{-0.3truecm}
\end{figure*}

Recently, \cite{cencini} have shown that to describe inertial pair
dispersion in synthetic flows it is useful to introduce the local or
{\it scale-dependent} Stokes number, using the ratio between the
particle response time and the typical eddy turnover time $\tau_R =
R/\delta_R u$ of the underlying fluid at a given scale: $St(R)=
{\tau_s}/{\tau_R} \sim \tau_s {\delta_R u}/{R}$. For real turbulent
flow where different scaling ranges are present, we can equivalently
define a scale dependent Stokes number $St(R)$ that recovers the usual
definition of the Stokes number $St(R) \simeq St = \tau_s/\tau_{\eta}$
when $R\ll\eta$ and behaves as $St(R)\sim
\tau_s\varepsilon^{1/3}R^{-2/3}$ when $R\gg\eta$. The typical
behaviour of $St(R)$ is sketched in Fig.~\ref{fig:stokes}, for two
different values of the Stokes number $St=3,70$ and using a
Batchelor-like parametrisation of the fluid velocity (see
\cite{meneveau}):
\begin{equation}
\label{eq:deltau}
\delta_R u \,=\,  U_0 \frac{R}{(\eta^2 + R^2)^{1/3}}.
\end{equation}

For Stokes numbers, $St$, order unity or larger, there always exists a
typical scale where the local Stokes number, $St(R)$, becomes order
unity,
\begin{equation} 
R^*(St) = \eta\, St^{3/2}.
\end{equation}
Such a scale, which is well in the inertial range if the Stokes number
$St$ is sufficiently large, can be considered a rough estimate of the
upper bound for the region of scales where inertia plays an important
role in the particle dynamics.
We expect that two main features might be important in characterising
the inertial particle stationary velocity statistics $\delta_R V $,
with respect to that of tracers $\delta_R u$. The first concerns the
small-scale behaviour of the particle velocity statistics. At small
scales $R \ll \eta$ and for large-enough Stokes numbers, the presence
of caustics makes the particle velocity increments not
differentiable. This feature can be accounted for by saying that
\begin{equation}
\label{eq:gamma_st}
  \delta_R V_{St} \sim
  V^{\eta}_{St}\left(\frac{R}{\eta}\right)^{\gamma(St)}; \qquad R \ll
  \eta,
\end{equation}
where the $V^{\eta}_{St}$ is a constant prefactor and the function
$\gamma(St)$ gives the typical scaling of caustic-like velocity
increments. Indeed we do expect that at changing the inertia of the
particles, the statistical weight of caustics might monotonically vary
as follows: at small $St$, $\lim_{St\rightarrow 0}\gamma(St) = 1$ ,
i.e.\ the value for smooth, differentiable Eulerian statistics of
tracers; at large values $St \rightarrow \infty$, it should approach
the discontinuous limit $\gamma(St) \rightarrow 0$, valid for
particles that do not feel underlying fluid fluctuations at all. The
right panel of Fig.~\ref{fig:stokes} shows the typical shape of the
function $\gamma(St)$ that is expected to be valid for turbulent
flows.

\begin{figure*}
\begin{center}
\includegraphics[width=1.1\textwidth]{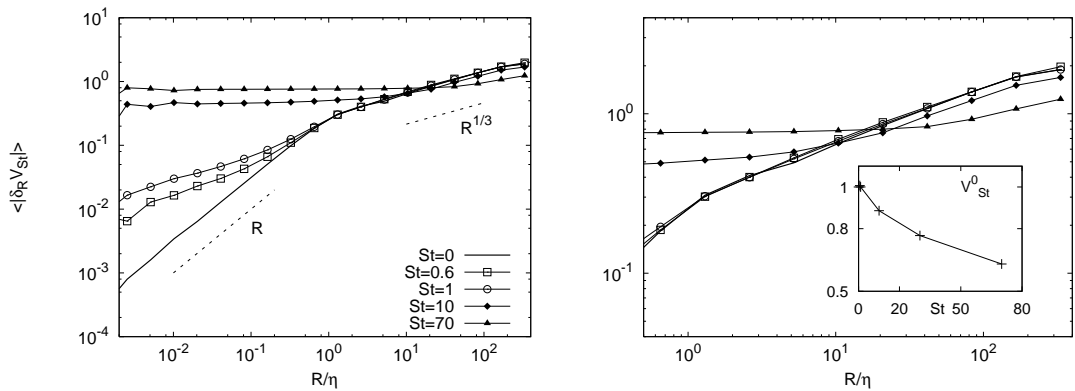}
\caption{Right figure: particle velocity structure function of order
  $p=1$ versus the scale $R/\eta$, for various Stokes numbers,
  $St=0,0.6,1,10,70$, and for Reynolds number $Re_{\lambda} ~\sim
  400$, Run II. The statistics for fluid tracers $(St=0)$ correspond
  to the solid line. Statistical errors are of the order of twice the
  size of symbols for scales smaller then $\eta$ and become comparable
  with the size of symbols in the inertial range of fluid velocity
  statistics. The differentiable scaling behaviour $\propto R$ in the
  dissipative range, and the Kolmogorov 1941 behaviour $\propto
  R^{1/3}$ in the inertial range of scales are also shown.  Left
  figure: zoom-in of the inertial range, same symbols as right.
  Inset: behaviour of the amplitude prefactor, $V^{0}_{St}$ as a
  function of the Stokes number $St$, as measured from the velocity
  increments at the integral scale $L$, Run II.}
\label{fig:1c}
\end{center}
\vspace{-0.3truecm}
\end{figure*}

The second important feature concerns the particle velocity statistics
at scales larger than the scale $R^*(St)$ previously defined, but
smaller than the integral scale of the fluid flow. For any fixed
Stokes number and for a large-enough Reynolds number, we expect that
inertia becomes weaker and weaker, by going to larger and larger
scales $R \gg R^*(St)$.  In such a case, particle velocity increments
are expected to approach the underlying fluid velocity increments:
\begin{equation}
  \delta_R V_{St} \rightarrow V^{0}_{St} \,\delta_R u \sim V^0_{St}
  \,U_0 \left(\frac{R}{L}\right)^{1/3}; \qquad R^*(St) \ll R \ll L,
\label{def:VtoU}
\end{equation}
where for simplicity we have neglected possible intermittent
correction to the Kolmogorov 1941 (K41) scaling of the fluid velocity
(see \cite{frisch} for details). Clearly, the Reynolds number has to
be sufficiently large to provide a well-developed scaling region
$R^*(St) \ll R \ll L$, before approaching the large scale $L$. We
emphasise that in~(\ref{def:VtoU}), an adimensional normalisation
factor $V^{0}_{St}$ has been introduced: it takes into account
possible filtering effects induced by inertia at large scales. The
normalisation is such that $V^{0}_{(St=0)}=1$, while for any Stokes
larger than zero $V^{0}_{(St)}\le 1$. \\ In Fig.~\ref{fig:1c} we test
the validity of the previous picture by analyzing the typical velocity
fluctuation, $\langle |\delta_R V_{St}|\rangle$, at changing Stokes
number and for data of Run II at Reynolds number $Re_{\lambda}\sim
400$. At small scales one detect the presence of caustics in the
velocity statistics, with a non smooth scaling behaviour below the
Kolmogorov scale $\eta$. At scales within the inertial range and when
the Stokes number is sufficiently large, the effect of caustics
affects also particle velocity statistics, up to a characteristic
scale which becomes larger and larger by increasing particle
inertia. Beyond this scale, particle velocity increments tend to
approach the scaling behaviour of the fluid tracers, but their
amplitude is depleted of a factor $1/V^{0}_{St}$, which increases with
the Stokes number, as shown in the inset of the right-hand panel of
Fig.~\ref{fig:1c}.
A similar behaviour is expected for higher order fluctuations, if we
neglect the role of intermittency.

It is interesting to consider the scaling behaviour of particle
velocity in terms of the underlying velocity statistics, not only at
very small or very large separations, but for any value of the scale
$R$. This is not straigthforward, since we have to account not only of
the fluid Eulerian statistics at the dissipative and inertial range of
scales, but also the modifications due to the inertia.  This is
responsible, as we have seen, for the appeareance of a new relevant
scale, and for filtering effects in the velocity amplitude.

To fully characterise particle velocity increments, we notice that the
Stokes scale, $R^*(St) $ defines a typical {\it Stokes-velocity}: this
is the fluid velocity increment at the Stokes scale, $\delta u^*(St)
\sim \delta_{R^*} u$ (see left panel of Fig.~\ref{fig:stokes}).
Previous reasonings can be summarised in the following interpolation
formula for the heavy particle velocity increment:
\begin{equation}
\label{eq:fit}
\delta_R V_{St}\, \,=\, \,V^{0}_{St} \, (\delta_R u)^{\gamma(St(R))}\,
\left[ (\delta_R u)^2+ c_1\, (\delta u^*(St))^2\right]^{ \left[
    1-\gamma(St(R))\right]/2}\,.
\end{equation}
The above expression is a Batchelor-like parametrisation but in the
velocity space, with a transient velocity given by the Stokes
velocity, $\delta u^*(St)$.

Once known the large scale normalization function $V^{0}_{St}$, the
caustic exponent $\gamma(x)$ (introduced in \cite{bccm05}) and the
reference fluid velocity increment $\delta_R u$, then the formula has
one free parameter only.  It is the prefactor $c_1$ appearing in front
of the Stokes velocity $\delta u^*(St)$, whose value depends again on
the inertia of the particles.
\begin{figure*}
\begin{center}
\includegraphics[width=0.6\textwidth]{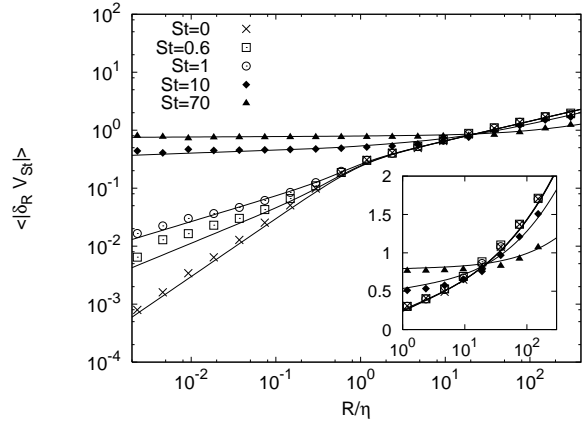}
\caption{Scaling behaviour of the particle velocity structure function
  of order one, versus the normalised scale $R/\eta$.  Solid lines:
  fit of the data of Fig.~\ref{fig:1c}, Run II, using the
  interpolation formula (\ref{eq:fit}). Here the large scale
  prefactors $V^{0}_{St}$ are those measured on Run II of the
  simulation, and shown in the inset of the right panel of
  Fig.~\ref{fig:1c}. Inset: enlargement of the crossover range, where
  $\delta_R u \sim \delta_R V$.}
\label{fig:3c}
\end{center}
\vspace{-0.3truecm}
\end{figure*}

In Fig.~\ref{fig:3c}, we show the result of the fit in terms of the
expression (\ref{eq:fit}), where the caustics scaling exponent has
been chosen as $\gamma(x) = \left[1 -2/\pi \, \mbox{atan}(x)\right]$:
this functional form provides a good fit to the numerical
results. Details of the small-scale caustic statistics will be
reported elsewhere.

The qualitative trend is very well captured by the interpolation
function proposed.  Notice that in (\ref{eq:fit}), the argument of
$\gamma(St)$ is not the simple Stokes number at the Kolmogorov scale,
but the scale-dependent one $St(R)$: $\gamma(St) \rightarrow
\gamma(St(R))$. This further ingredient is needed to take into account
the fact that in presence of a rough underlying fluid velocity, as it
happens in the inertial range of scales, no simple power law behaviour
is expected for the scaling of particle velocity statistics.  This was
previously remarked in \cite{cencini}, in the study of heavy particle
turbulent dispersion in random flows.

Equation (\ref{eq:fit}) clearly matches the two limiting behaviours
for very small and very large separations. In the former case, inertia
dominates the small-scale velocity statistics with respect to the
underlying smooth fluid velocity, and caustics lead to a pure
power-law behaviour,
\begin{equation}
\label{caseA}
\delta_R V \sim (\delta_R u)^{\gamma(St)} \sim V^{0}_{St}
\,\left(\frac{R}{L}\right)^{\gamma(St)}; \qquad R \ll \eta,
\end{equation}
where the local Stokes number has attained its dissipative limit
$St(R) \rightarrow St$.

In the latter case, at very large scales $R \gg R^*(St)$ inertia is
subleading, and the typical velocity difference between particles is
close to the fluid velocity increment,
\begin{equation}
\label{caseC}
\delta_R V_{St} \sim V^{0}_{St} \,\delta_R u; \qquad \eta \ll R^*(St) \ll R.
\end{equation}
At intermediate scales, for large Stokes, $St \ge 1$, inertia brings a
non-trivial dependency via the scale-dependent Stokes number, $St(R)$,
and we expect a pseudo power-law scaling:
\begin{equation}
\label{caseB}
\delta_R V_{St} \sim (\delta_R u)^{\gamma(St(R))} \sim R^{\gamma(St(R))/3};
\qquad  \eta \ll R \ll R^*(St)\,.
\end{equation}

\begin{figure*}
\begin{center}
\includegraphics[width=0.7\textwidth]{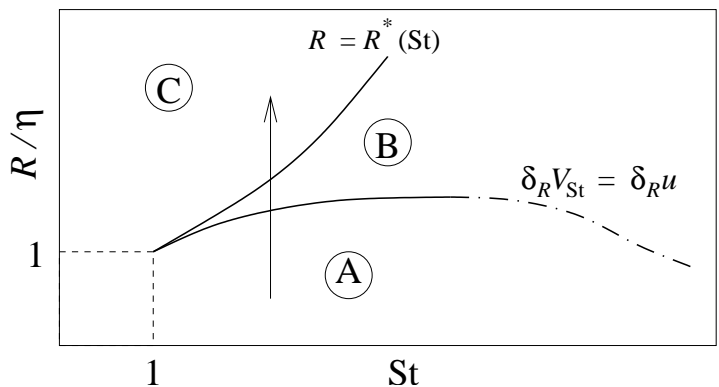}
\caption{Sketch of the different regimes expected in the parameter
  space of inertia, $St$, and scale separation $R$. The curve
  $St(R^*)=1$ separate the region of low inertia $St(R) \le 1$, region
  (C), from the regions where inertia is important $St(R) \ge 1$,
  regions (A) and (B). Further we can distinguish the separation
  regime where inertia is important and particle velocity difference
  larger than the fluid one at the same scale, region (A), from the
  intermediate regime where inertia is still important particle
  velocity difference is smaller that the corresponding fluid one,
  region (B). Separation between region (A) and (B) is given by the
  curve $\delta_R V_{St}= \delta_R u$. For relative dispersion of
  pairs of Stokes $St$ starting at a given separation $R$, one
  typically starts from the corresponding position in this plane and
  then evolve upwards along the vertical arrow.}
\label{fig:abc}
\end{center}
\vspace{-0.3truecm}
\end{figure*}

Summarising, we propose that at changing the Stokes and Reynolds
numbers, different regimes governing the particle velocity statistics
can be distinguished.  The relevance of such regimes of the particle
velocity statistics for the associate relative dispersion dynamics can
be easily explained with the help of the sketch reported in
Fig.~\ref{fig:abc}. In the parameter space of inertia and scale
separation $(St, R)$, we can distinguish three regions depending
whether inertia is strong or weak, and whether particle velocity
difference is large or not with respect of the fluid velocity
difference at comparable scale.  In agreement with what commented
before, we pose that the curve $St(R^*)=1$ distinguishes the region of
weak ($St(R) \le 1$) and strong inertia ($St(R) \ge 1$).

Regime (A) is such that inertia is important since the scale $R^*(St)
\gg \eta$, and moreover the typical particle velocity increments are
larger than the fluid increments.  In the region (B), inertia is still
important but particle velocity increments are depleted with respect
to the fluid increments. This typically happens for large Stokes
numbers, and in our DNS is visible only for very large separations
$R(t)$ of the highest Stokes $St=70$. Finally, regime (C) is
characterised by a weak inertia, which appears only in the filtering
factor for the velocity large-scale amplitude and possibly in
sub-leading corrections to the tracers relative dispersion.

Even for the largest value of the Reynolds and Stokes numbers achieved
in our DNS, it is very difficult to disentangle quantitatively the
above mentioned regimes, because of the closeness of the three
relevant scales, $\eta$, $R^*(St)$, and $L$.
Still, the quality of the fit shown in Fig.~\ref{fig:3c} using the
global functional dependence given by Eqn. (\ref{eq:fit}) makes us
confident that the main physical features are correctly
captured. Before closing this section, we note that there is no reason
to assume that the functional form entering in the pseudo-power law
scaling in the inertial range, $\gamma(St(R))$, in (\ref{caseB}) is
equal the one characterizing the scaling in the viscous range,
$\gamma(St)$, in Eqn. (\ref{caseA}).  Hint for this observation come
from results obtained in \cite{cencini} for random flows, where a very
high statistical accuracy can be achieved: there, depending if the
underlying fluid velocity is spatially smooth or rough, a slightly
different functional form has been found.

The previous analysis gives us a clear quantitative picture of the
scale and velocity ranges where caustics play a role in the particle
dynamics. For example, for moderate Stokes numbers, we have important
departure from the tracers statistics only for very small scales,
i.e.\ caustics gives a singular contributions to the particle velocity
increments inside the viscous range; then, at larger scales, the
particle velocity scaling become indistinguishable from the tracer
velocities. Clearly, for such Stokes, no important corrections for
particle separation evolution is expected with respect to the usual
Richardson dispersion observed for tracers. This is because particle
pairs tends to separate, and very soon all pairs will attain
separations where their velocities are very close to the underlying
fluid. On the other hand, for very heavy particles, those with Stokes
time falling inside the inertial range of fluid velocity statistics,
the contribution from the caustics will be felt also at relatively
large scales, up to $ R \sim R^*(St)$. Pair separations attain such
scales when the initially large relative velocity difference has
relaxed and become smaller than the corresponding fluid
one\,---\,crossing from region (A) to (B). Notice that at $R \sim
R^*(St)$, we have that $\delta_{R^*}V_{St} \simeq V^0_{St}
\delta_{R^*} u$, i.e.\ there is a non-trivial effect from inertia.
Moreover, for large Stokes, at scales $R > R^*(St)$, particle velocity
increments are smaller than the fluid counterparts, indicating an
important depletion induced by the Stokes drag on the particle
evolution.\\ It is clear from the above discussion that new physics
should appear for the value of inertia and scales separation of region
(B). This regime\,---\,that we can not access with the present
data\,---\,is the one where a new law of pair separation should appear
as recently suggested by \cite{fouxon}. A discussion of the dispersion
regimes of inertial particle pairs follows in the next section in
terms of the time behaviour of the mean square separation distance.

\section{Dispersion regimes and corrections due to inertia}
\label{sec:2}
In this section we analyse the effects of inertia on the mean square
separation of heavy particle pairs with a given initial separation
distance, $R_0$ at time $t=t_0$, as a function of the Stokes number:
\begin{equation}
\langle (R(t))^2 \,|\, R_0,t_0 \rangle_{St} \,=\, \langle |\bm X_1(t)
-\bm X_2(t)|^2 \rangle_{St},
\label{eq:disp}
\end{equation}
where in the left-hand side the average is performed over all pairs of
particles such that $|\bm X_1(t_0) - \bm X_2(t_0)| = R_0$.  The study
of the relative dispersion of small, neutrally buoyant tracer
particles has recently been the subject of renewed interest. This has
been motivated by the fact that very accurate\,---\,highly resolved in
time and space\,---\,data have become available, experimentally
(\cite{OM,bodi.science}) and numerically (\cite{YB,bif.2p,noi.jot}).
These studies have confirmed what was known since the works of
\cite{rich.rev} and \cite{batch.rev}, i.e.\ the existence of different
dispersive regimes for tracer pairs in turbulent flows, depending on
the value of their initial distance and on the time scale considered.

When released in a statistically homogeneous and isotropic, turbulent
flow with an initial separation $R_0$ in the inertial range for fluid
velocity, i.e.\ $\eta \simeq R_0 \ll L$, tracer pairs initially
separate according so the so-called Batchelor regime,
\begin{equation} 
  \langle (R(t))^2 | R_0,t_0 \rangle_{St=0} \simeq R_0^2 + C
  (\varepsilon R_0)^{2/3}\,t^2; \qquad \tau_{\eta} \ll (t-t_0) \ll
  t_B\,,
\label{eq:batch}
\end{equation}
where $C$ is supposed to be a universal constant, and $\varepsilon$ is
the average kinetic energy dissipation of the flow.  This {\it
  ballistic} regime appears because initially tracers separate {\it as
  if} the underlying velocity field were frozen, and it lasts for a
time scale that is a function of the initial separation itself, $t_B=
\left(R_0^2/\varepsilon\right)^{1/3}$ (see
\cite{batch.rev,bodi.science}).  After such a transient initial time,
the relative separation dynamics forgets the initial conditions and
tracers separate explosively with a power law behaviour given by the
Richardson law:
\begin{equation}
\langle (R(t))^2
| R_0, t_0 \rangle_{(St=0)} \sim g\, t^3; \qquad t_{B} \ll (t-t_0) \ll
T_L\,,
 \label{eq:r2rich} 
\end{equation}
where $g$ is known as the Richardson constant. As set out in Monin \&
Yaglom \cite{MY}, the tracer separation PDF\,---\,that will be
discussed later\,---\,has a similar scaling behaviour in these ranges.

A remarkable fact of the Richardson dispersion (\ref{eq:r2rich}) is
the disappearance of the dependence on the initial separation $R_0$,
an effect also dubbed \emph{intrinsic stochasticity} (\cite{eve00}),
which is just the signature of the non-Lipschitz nature of the
velocity field driving the separation between tracers, when their
mutual distance is in the inertial range of fluid velocity statistics.
The experimental and numerical validation of the previous prediction
(\ref{eq:r2rich}) has proved to be particularly difficult, the main
reason being the strong contamination from viscous and large scale
effects in the tracers dynamics. To overcome these problems, a series
of techniques have been developed, including the study of {\it
  doubling time statistics}; i.e.\ the probability distribution
function of the time needed for a pair to double its separation
(\cite{boffi.2d,bif.2p}). Thanks to these techniques, a fairly good
agreement on the value of the Richardson constant has been achieved.
\begin{figure*}
\begin{center}
\includegraphics[width=1.\textwidth]{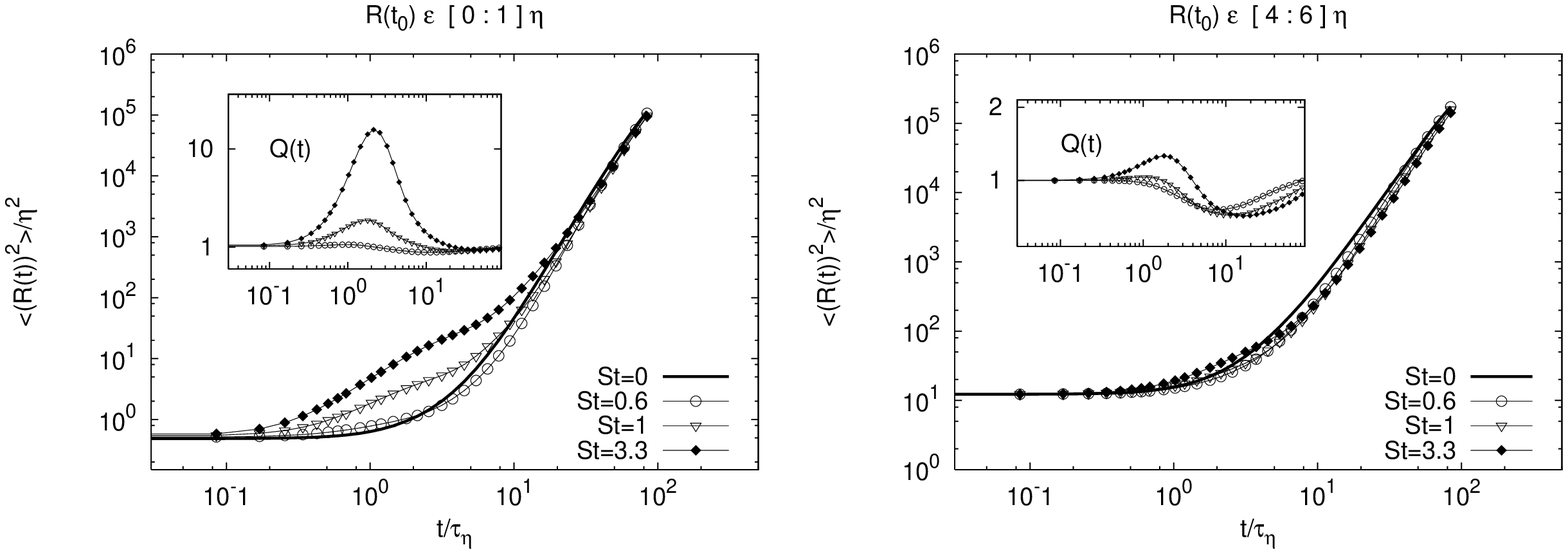}
\caption{Mean square separation versus time, for heavy particles at
  changing $St$ and the initial distance $R_0$. Time is normalised
  with the Kolmogorov time scale $\tau_{\eta}$. Left panel:
  $St=0,0.6,1,$ and $3.3$; initial distance $R_0 \in [0:1]\eta$, Run
  I. Error bars due to statistical fluctuations are of the order of
  the symbol size. Notice that the two largest Stokes numbers show a
  time lag interval where separation proceeds faster than
  tracers. Inset: ratio between the heavy particle separation and the
  tracer data, $Q(t)$ versus time, for $St=0.6,1,$ and $3.3$. Same
  symbols as in the body of the figure are used. Right panel: mean
  square separation versus time, but with a larger initial distance,
  $R_0 \in [4:6]\eta$. Stokes numbers are the same as in the left
  panel. Notice that now only the dispersion of particle pairs with
  $St=3.3$ exhibits a small departure from the underlying fluid, as
  shown by the $Q(t)$ indicator in the inset. For the smaller Stokes,
  typical size of caustics is smaller than the initial separation
  $R_0$, and particle pairs therefore separate as fluid tracers do.}
\label{fig:m2}
\end{center}
\vspace{-0.3truecm}
\end{figure*}
Here, we want to study how the tracer behaviour is modified by the
presence of small-scale caustics in particular and by inertia effects
in general, for the case of heavy particle pairs.  Standard direct
measurements of the moments of separation as a function of time will
be considered, while application of doubling time statistics is left
for future studies.

In Fig.~\ref{fig:m2}, we show the behaviour for the mean square
separation at varying the Stokes number, and for two values of the
initial separation.  We start with data at the lowest resolution,
i.e.\ Run I at $Re_{\lambda} \simeq 200$, and for moderate Stokes
numbers, $St \sim {\cal O}(1)$. Initial distances are chosen equal to
$R_0 \le \eta$ (left panel) and $R_0 \in [4\!:\!6]\,\eta$ (right
panel).

If the initial distance is small enough (left panel), the presence of
caustics in the particle velocity field at initial time gives a very
remarkable departure from the tracer behaviour. At increasing the
Stokes number, such departure is more and more evident, and it lasts
for a time lag which becomes longer and longer.  For the highest value
of the Stokes number shown in the left panel of Fig.~\ref{fig:m2}
($St=3.3$), a sensible difference from the tracer behaviour is
observed over almost two decades: $t \in [0.1\!:\!10]\,\tau_{\eta}$.
A way to better visualise the departure from the tracer statistics
consists in plotting the mean square separation for heavy pairs of
different Stokes numbers, normalised to the tracer one, that is
\begin{equation}
  Q(t) \,=\, \frac{\langle (R(t))^2\rangle_{St}}{ \langle (R(t))^2
    \rangle_{(St=0)}}.
  \label{eq:defQ}
\end{equation}
This quantity is shown in the insets of Figure \ref{fig:m2}.  For
heavy pairs starting at $R_0 \simeq \eta$ and with $St=3.3$, the
relative difference is as large as $10$ at its maximum for $t \sim
\tau_{\eta}$. However, such effect becomes progressively less
important if we start the separation experiment from larger initial
distances as shown in the right panel of the same figure.  This is
because, at these same Stokes numbers, the deviation of particle
velocity difference with respect to the underlying fluid, due to
caustics, has already decreased. This is equivalent to state that, for
these Stokes numbers, the typical size of caustics is smaller than the
initial separation $R_0$, and particle pairs therefore separate as
fluid tracers do. At larger time lags, whatever the value of the
initial separation, the Richardson dispersion regime is recovered. \\
\begin{figure*}
\begin{center}
\includegraphics[width=1.\textwidth]{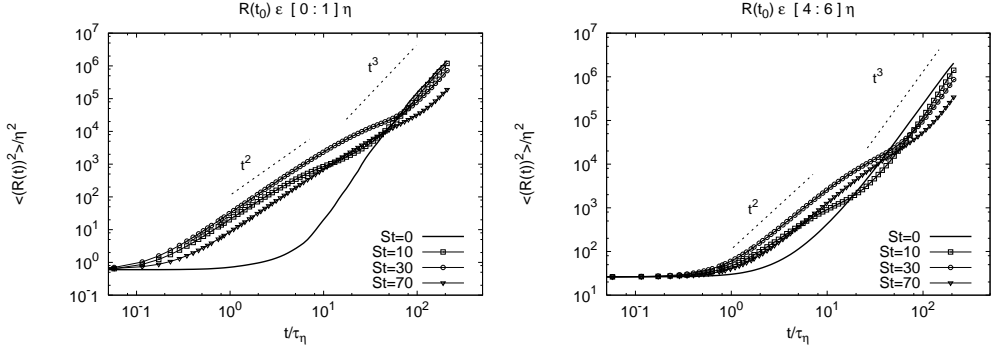}
\caption{Mean square separations versus time, for pairs with $St=10,
  30, 70$, at $Re_{\lambda}= 400$. Left and right panels refer to the
  two initial distance $R_0 \in [0\!:\!1]\eta$ and $R_0 \in
  [4\!:\!6]\eta$, respectively. Error bars due to statistical
  fluctuations are of the order of the symbols size. Tracers (solid
  lines) are also shown for comparison. Notice the ballistic behaviour
  for the heavy particle separation observed in the {\it caustics
    dominated} time interval. For very large time-lags a
  Richardson-like behaviour starts to develop but with a less intense
  overall speed of separation, due to the depletion effects of the
  $V^{(0)}_{St}$ prefactor in the particle velocity increments for
  large Stokes numbers.  The slopes of the Batchelor, $\langle R^2(t)
  \rangle \propto t^2$, and Richardson, $\langle R^2(t) \rangle
  \propto t^3$ dispersion regimes are also drawn for reference.}
    \label{fig:st70}
  \end{center}
  \vspace{-0.3truecm}
\end{figure*}
We now consider what happens for larger Stokes numbers. In
Fig.~\ref{fig:st70}, we show the results for the mean square
separation of $St=10, 30,$ and $70$ and for the large Reynolds number,
$Re_{\lambda} \simeq 400$.  Both initial distances, $R_0 \in [0\!:\!1]
\eta$ and $R_0 \in [4\!:\!6]\eta$ are displayed.  As one can see, for
the large value of $St=70$, the tracer-like behaviour is
never recovered, and even the separation of pairs starting with the
largest distance $R_0$ is affected. The transient regime dominated by
the caustics invades the whole inertial range: since particle pairs
need a very long time to decrease their initial velocity difference to
the value of the fluid increment at the corresponding scale, they
separate with a quasi-ballistic behaviour:$ \langle R^2(t)
\rangle_{St} \propto t^2$.\\ The above scenario can be interpreted in
terms of caustic-{\it dimensions}. At any value of the inertia, there
exist a spatial length, of the order of the scale $R^*(St)$, which
identifies the typical spatial size of caustics, i.e.\ the range of
scales where particle velocity increments are uncorrelated from the
underlying fluid velocity field. If the initial pair separation $R_0$
is taken inside this region (left panel of Fig.~\ref{fig:m2}),
particle pair separation starts much faster than for fluid tracers,
because of the much more intense velocity differences felt by the
pairs inside the caustics. When particle pairs reach a separation
larger than $R^*(t)$, they start to be synchronised with the
underlying fluid velocity, recovering the typical Richardson
dispersion. However, if the initial separation is larger than the
caustics size, the evolution of inertial particle pairs is almost
indistinguishable from the tracers. Finally, whether or not a
Richardson-like behaviour is recovered for very large inertia, may
depend on the Reynolds number also. In the limit of larger and larger
Reynolds, at fixed Stokes number, one may expect a final recovery of
the fluid tracers behaviour even for very heavy particle pairs.

\subsection{Mean-field approach to heavy particle dispersion}
\label{subsec:mean}
The turbulent relative dispersion of fluid tracers can be easily
modelled by applying K41 scaling theory to the fluid
velocity increments governing particle separation dynamics (see,
e.g.,~\cite{bodi.njp}). Indeed, if ${\bm R}(t)$ is the tracer
separation vector at a given time, its evolution is completely
specified by the equation
\begin{equation}
  \dot{\bm R}(t)\, =\, {\bm u}({\bm X}_1,t) - {\bm u}({\bm
    X}_2,t)\,=\, \delta_R \bm u ({\bm R},t) \,,
\label{eq:tracdisp}
\end{equation}
together with the initial condition ${\bm R}(t_0)={\bm R}_0$. Hence,
we can directly write an equation for the root-mean-square separation
$r(t) \,\equiv\, \langle |{\bm R}(t)|^2 \,|\, R_0, t_0 \rangle^{1/2}$
\begin{equation}
  \dot{r} = \frac{1}{r}\left\langle \bm R(t) \cdot \delta_R \bm u
  ({\bm R}(t),t) \,|\, R_0, t_0 \right\rangle, \quad\mbox{with }
  r(t_0) = R_0.
\end{equation}
We next assume the following mean-field closure for the right-hand
side:
\begin{equation}
  \langle \bm R(t) \cdot \delta_R \bm u(\bm R(t),t) \,|\, R_0, t_0
  \rangle \approx \langle \bm R^2 \,|\, R_0, t_0 \rangle^{1/2} \,
  \langle \hat{\bm R}\cdot\delta_R \bm u \rangle \,=\, r\,
  S_1^{\mbox{\tiny/\!/}}(r), \label{eq:c1}
\end{equation}
where $S_1^{\mbox{\tiny/\!/}}(r)$ is the first-order Eulerian
longitudinal structure function of the underlying homogeneous and
isotropic turbulent flow. According to K41 phenomenology, this
structure function behaves in the inertial range as
$S_1^{\mbox{\tiny/\!/}}(r)\simeq C\,\varepsilon^{1/3}r^{1/3}$, where
$C$ is an order-unity constant. This closure finally leads to
\begin{equation}
  \dot{r} = C\,\varepsilon^{1/3}r^{1/3},\quad\mbox{so that}\quad r(t)
  = \left[ R_0^{2/3} + (2C/3)\, \varepsilon^{1/3} (t-t_0) \right]^{3/2}.
  \label{eq:mf-tracers}
\end{equation}
Such an approximation gives a complete qualitative picture of the time
evolution of the mean square separation between tracers. In
particular, it encompasses the two important regimes of relative
dispersion: when $(t-t_0)\ll t_B = (3/2C)\,\varepsilon^{-1/3}
R_0^{2/3}$, a Taylor expansion of the solution (\ref{eq:mf-tracers})
gives the Batchelor regime $r(t)\simeq R_0 + C\,(\varepsilon
R_0)^{1/3}\,(t-t_0)$, while when $(t-t_0)\gg t_B$, one recovers
Richardson's law $r(t)\simeq (2C/3)^{3/2}\varepsilon^{1/2}\,t^{3/2}$.

In the case of inertial particles, the number of degrees of freedom to
describe the dynamics is obviously increased: the separation between
two heavy particles obeys
\begin{equation}
  \ddot{\bm R}(t) \, =\, - \frac{1}{\tau_s}\left[\dot{\bm R}(t) -
    \delta_R \bm u ({\bm R},t) \right].
\label{eq:heavydisp}
\end{equation}
In order to derive {\it mean-field} equations one has to track
simultaneously the average distance and velocity difference between
particles. For this we follow the same spirit as for tracers and
introduce the particle velocity structure function $v(t) \,\equiv\,
\langle |\delta_R {\bm V}(t)|^2 \,|\, R_0, t_0 \rangle^{1/2}$, where
$\delta_R \bm V(t) \,=\, \dot{\bm R} (t)$ is the velocity difference
between the two particles. One can proceed as previously to write from
(\ref{eq:heavydisp}) exact equations for $r(t)$ and $v(t)$:
\begin{eqnarray}
  \ddot r \,&=&\, \frac{1}{r} (v^2 - \dot r^2) - \frac{1}{\tau_s}
  \left[ \dot r - \frac{1}{r}\,\langle \bm R \cdot \delta_R \bm u \rangle
    \right], \label{eq:rexact}\\ \dot v \,&=&\, -
  \frac{1}{\tau_s}\left[ v - \frac{1}{v}\, \langle \delta_R \bm V \cdot
    \delta_R \bm u \rangle \right],\label{eq:vexact}
\end{eqnarray}
where for the sake of a lighter notation the indication of conditional
ensemble averages was dropped.  It is worth noticing that the
root-mean-square velocity difference $v(t)$ evolves with a dynamics
that resembles closely that of heavy particles. However, $v(t)$ does
not coincide with the time derivative of the mean distance $r(t)$. It
is thus useful to rewrite the above equations introducing a sort of
transverse particle velocity component $w$ defined as
\begin{equation}
  \dot r \,=\, v-w\,. \label{eq:v-w}
\end{equation}
We can write an exact equation also for the evolution of $w$
\begin{equation}
  \dot w \,=\, -\frac{1}{\tau_s} w - (2v-w)\frac{w}{r} -
  \frac{1}{\tau_s} \left[\frac{1}{r} \langle \bm R \cdot \delta_R \bm u
    \rangle - \frac{1}{v} \langle \delta_R \bm V \cdot \delta_R \bm u
    \rangle \right]\,. \label{eq:wexact}
\end{equation}
Of course, equations (\ref{eq:vexact}), (\ref{eq:v-w}), and
(\ref{eq:wexact}) are not closed without supplying the correlation
between the particle evolution and the underlying fluid. As in the
case of tracers, the first unclosed term appearing in the right-hand
side of (\ref{eq:wexact}) is approximated by (\ref{eq:c1}). The next
unclosed term involving the correlation between fluid and particle
velocity differences is approximated by
\begin{equation}
  \langle \delta_R \bm V \cdot \delta_R \bm u \rangle \approx \langle
  |\delta_R \bm V|^2\rangle^{1/2} \, \langle |\delta_R \bm u|^2
  \rangle^{1/2} \,=\, v \, S_2^{1/2}(r) \,,\label{eq:c2}
\end{equation}
where $S_2(r)$ denotes the full second-order structure function of the
fluid velocity field. When $r$ is in the inertial range, K41
phenomenology implies that $S_2(r) \propto (\varepsilon
r)^{2/3}$. Finally these approximations lead to a closed set of
equations for the time evolution of the average separation and
velocities $r$, $v$, and $w$
\begin{eqnarray}
  \dot r \,&=&\, v-w\,,\label{eq:mf1}\\ \dot v \,&=&\,
  \frac{1}{\tau_s}[C\,\varepsilon^{1/3}r^{1/3} -v]
  \,,\label{eq:mf2}\\ \dot w \,&=&\, -\frac{1}{\tau_s}w
  -(2v-w)\,\frac{w}{r}+ \frac{1}{\tau_s} B\,
  \varepsilon^{1/3}r^{1/3}\,, \label{eq:mf3}
\end{eqnarray}
where $B$ and $C$ are positive order-unity dimensionless constants,
reflecting the lack of control on the prefactors of the scaling laws
in the closures (\ref{eq:c1}) and (\ref{eq:c2}). This system of
equations is supplemented by the initial conditions $r(t_0)=R_0$,
$v(t_0) = \langle |\delta_{R_0} \bm V |^2\rangle^{1/2}$ and $w(t_0)=
v(t_0) - \langle \bm R_0 \cdot \delta_{R_0} \bm V \rangle / R_0$,
which clearly depend on the dispersion experiment under
consideration. It is worth noticing that this system of equations
reduces to the mean-field equation (\ref{eq:mf-tracers}) for
tracers in the limit of vanishing inertia $\tau_s\to 0$.

\begin{figure*}
\begin{center}
\includegraphics[width=0.52\textwidth]{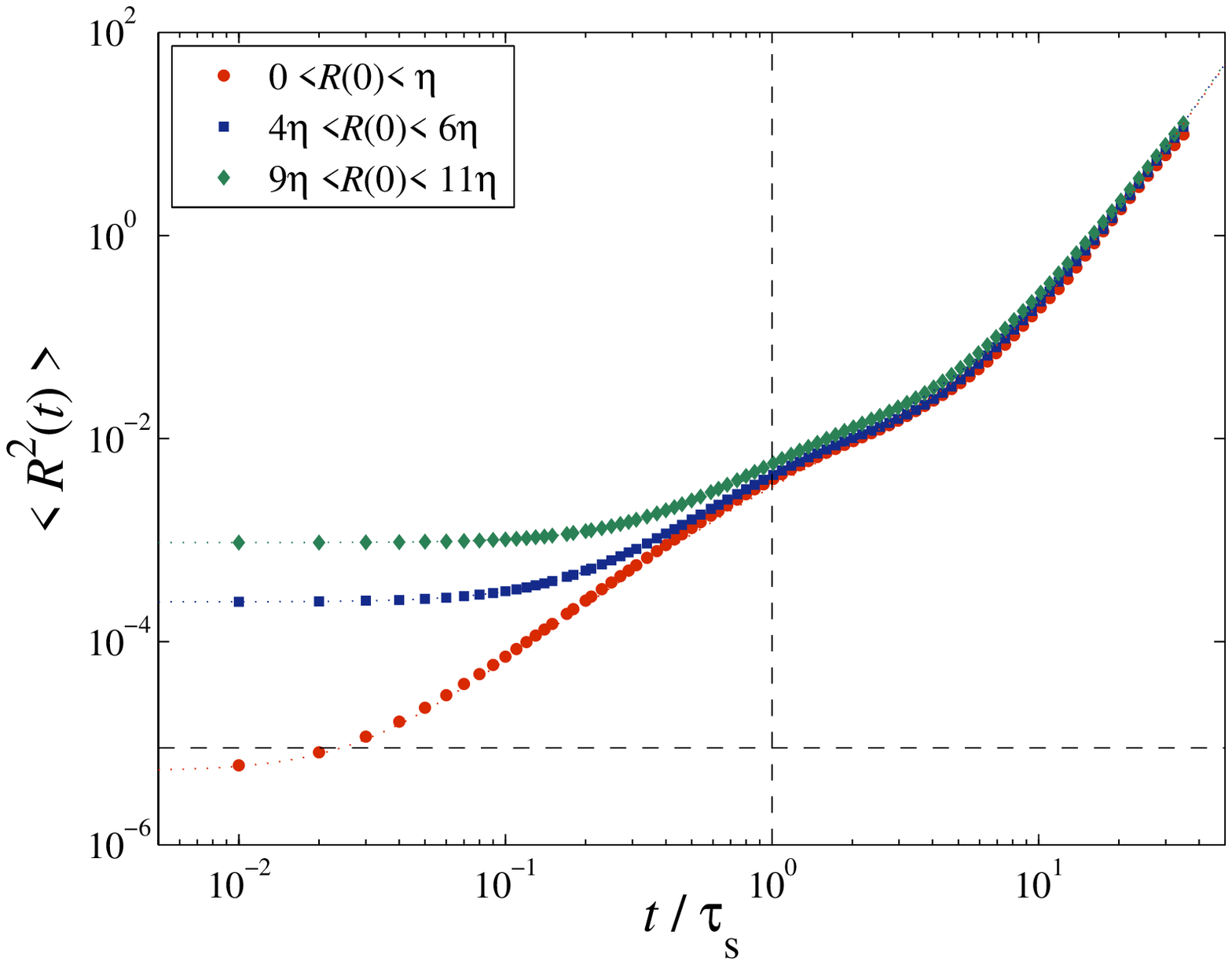}\hspace{-20pt}
\includegraphics[width=0.52\textwidth]{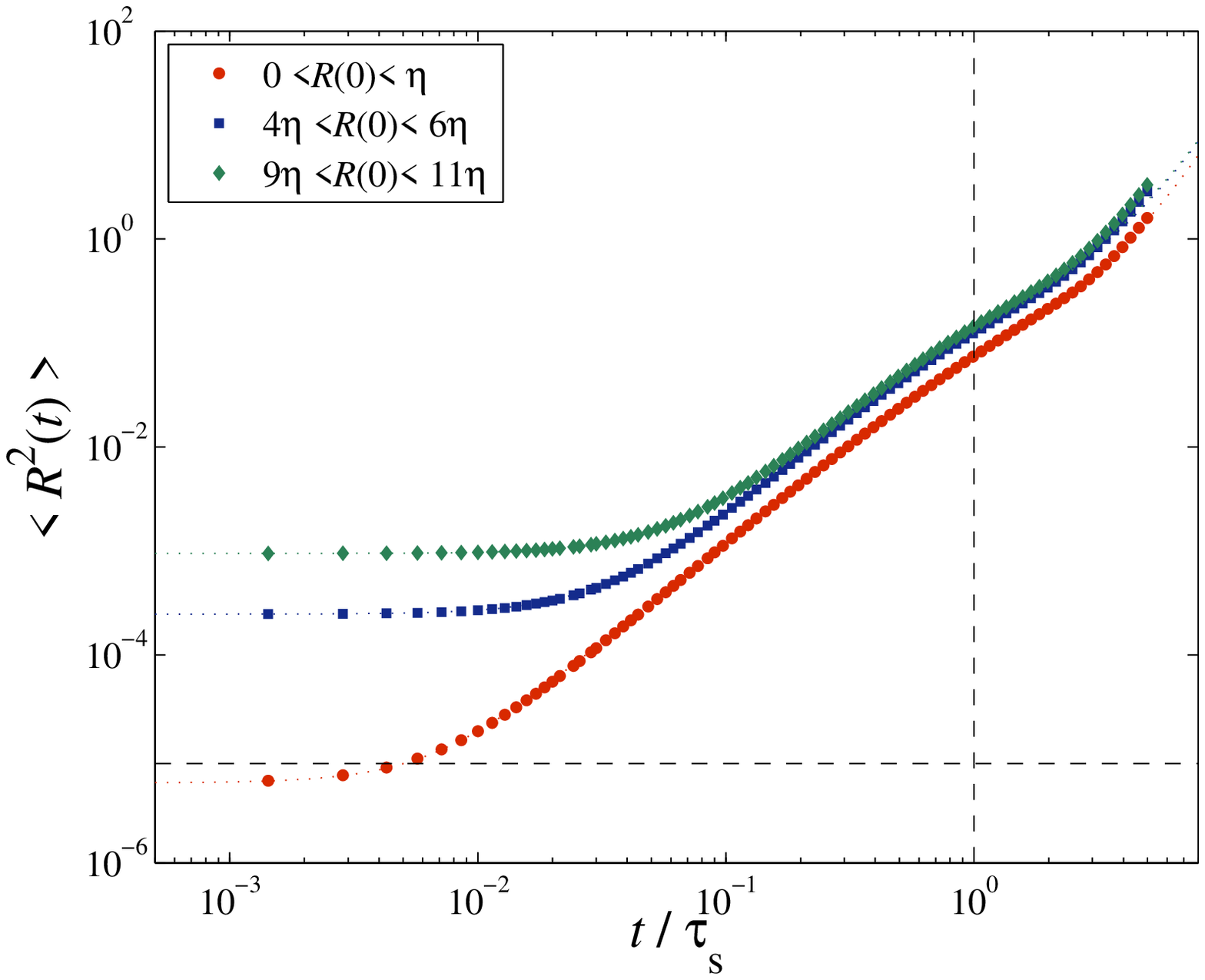}
\caption{Mean square separation versus time, for Run II and Stokes
  numbers $St=10$ (left panel) and $St=70$ (right panel). Data are
  shown for the available choices of the initial separations : $R_0
  \le \eta$, $R_0 \in [4\!:\!6]\,\eta$, and $R_0 \in [9\!:\!11]\,\eta$
  (from bottom to top). Symbols stand for DNS data, while solid lines
  are mean-field solutions. Adimensional prefactors are $B=0$ and
  $C\simeq2$. Notice that time is made adimensional with the Stokes
  time $\tau_s$. Vertical dashed lines mark $t/\tau_s=1$.}
\label{fig:jeremie1-2}
\end{center}
\vspace{-0.3truecm}
\end{figure*}

Similarly to the case of tracers, the crude approximation
(\ref{eq:mf1})-(\ref{eq:mf3}) of the evolution of the root-mean-square
distance between heavy inertial particles is able to capture the main
features of the separation time behaviour. In
Fig.~\ref{fig:jeremie1-2}, we show the result of the numerical
integration of the set of equation (\ref{eq:mf1})-(\ref{eq:mf3})
obtained by an appropriate choice of the free parameters (see figure
caption), together with DNS data from Run II, for two different large
values of the Stokes number.
For fixed initial separation and at increasing the intensity of the
caustics velocity increments in the initial condition (i.e.\ at
increasing inertia), the transient deviation from the Richardson
behaviour become more and more evident at intermediate times (of the
order of the Stokes time $\tau_s$, not shown). Clearly such a simple
approach can be valid only in a limited region of the phase space,
where the initial conditions are, at least, at the edge of the
inertial range so that K41 scaling is correct for the fluid velocity
second-order increments. Moreover, the matching scale where particle
velocity increments become of the order of the fluid increments has to
fall in the inertial range too: if this is not the case, then pairs enter
the regime where inertia is important but particle relative velocity
is small at scales of the dissipative range, and the mean-field
closure proposed above becomes inadequate. A detailed quantitative
comparison of the realm of applicability of the mean-field
approach\,---\,including effects of viscous scales and small-scale
caustics\,---\, will be the object of future work.

\subsection{Cross-over between relaxation to the fluid velocity and
  Richardson behaviour }
\label{sub:crossover}
Despite its simplicity, the mean-field approach described above is
able to correctly reproduce the pair dispersion of heavy particles
with initial data in the inertial range. We might wonder if one can
draw an even simpler qualitative picture of pair dispersion. For this,
we consider the behaviour of particle pairs with moderately large
Stokes numbers, for which inertia plays an important role for the
initial transient and the Richardson behaviour is slowly recovered
well inside the inertial range of scales. For simplicity, we assume
that the scale where the fluid and particle velocity becomes of the
same order, $\delta_R \bm V \sim \delta_R\bm u$, and the scale
$R^*(St)$, where inertia ceases to be important, are very close. As it
is clear from the sketch of Fig.~\ref{fig:abc}, this may not be always
the case because of the effect of the normalisation factor
$V^{0}_{St}$ for large Stokes: in the picture, it corresponds to
Stokes number with a narrow transient region (B).

\begin{figure*}
\begin{center}
\includegraphics[width=0.52\textwidth]{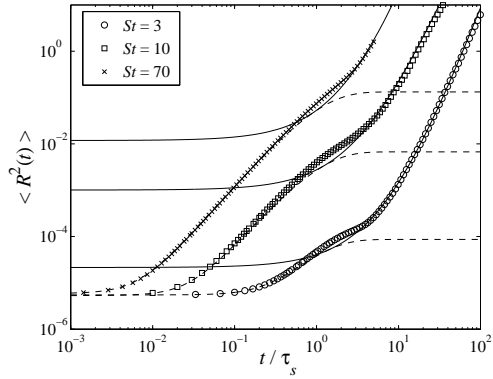}
\caption{Mean square separation versus time from DNS data of Run II,
  for three different values of the Stokes numbers. Notice that time
  is normalised with the Stokes time $\tau_s$. Solid lines represent the
  initial, almost ballistic evolution due to the exponential
  relaxation of velocity statistics, and the dashed lines correspond
  to the Richardson regime. With a suitable tuning of the free
  parameters, here the Richardson constant $g$ and the initial
  velocity increment value $\langle (\delta_R V(t_0))^2\rangle$, both
  temporal behaviours are reproduced.}
\label{fig:jeremie3}
\end{center}
\vspace{-0.3truecm}
\end{figure*}

The general picture then goes as follows. Initially particles separate
almost ballistically during a time which is of the order of (or larger
than) the time needed by their initial, caustics-dominated, velocities
to relax to the fluid velocity. After that time, particles behave as
tracers and reconcile with a standard Richardson dispersion. This is a
first order approximation since (\textit{i}) the fluid flow actually
correlates to the particle dynamics already at very small times, and
(\textit{ii}) inertia effects are present up to large times as
previously discussed. Nevertheless, such an approximation should give
the two correct qualitative asymptotic behaviours, at small and large
time scales. Since we consider moderately large values of the Stokes
number, the initial typical particle velocity can be assumed to be
much larger than the fluid velocity, i.e.\ $|\delta_R \bm V|\gg
|\delta_R\bm u|$. Under these hypotheses, there is an initial time
interval during which difference between particle velocities obeys
$\delta_R \dot{\bm V} \approx -(\delta_R \bm V)/\tau_s$ [see
(\ref{eq:vexact})], and thus $\delta_R \bm V(t) \simeq (\delta_R \bm
V(t_0))\,e^{-(t-t_0)/\tau_s}$. As a consequence, the mean square
separation between particles evolves initially as:
\begin{eqnarray} 
\langle |\bm R^2(t)| \,|\, R_0,t_0 \rangle &=& R_0^2 + 2 \tau_s
\langle \bm R(t_0) \cdot \delta_R \bm V(t_0) \rangle
(1-e^{-(t-t_0)/\tau_s}) \nonumber \\ &&+ \, \tau_s^2 \langle (\delta_R
\bm V(t_0))^2\rangle (1-e^{-(t-t_0)/\tau_s})^2.
\label{eq:crossR} 
\end{eqnarray}
This should be approximately valid up to a time scale, in the inertial
range, where $|\delta_R \bm V| \sim |\delta_R \bm u| \sim (\varepsilon
R)^{1/3}$: it is easy to show that such a time scale is proportional
to the particle response time $\tau_s$. For larger times, inertia
effects become subdominant and heavy pair dispersion suddenly gets
synchronised to a Richardson like regime. Nevertheless, this
Richardson regime has started only after the previous relaxation has
ended, that is at a distance much larger than the original separation
$R_0$ of the particle pair. The combination of this initial
exponential relaxation of heavy particles with moderately large
inertia, plus the later standard Richardson diffusion are the two main
features due to inertia in the inertial pair dispersion. This is
indeed confirmed by Figure~(\ref{fig:jeremie3}), where we compare DNS
data for mean square separation, with the two phenomenological regimes
just described, for which we have assumed that $\langle \bm R \cdot
\delta_R \bm V(t_0) \rangle \simeq 0$.  As we can see, the main
qualitative trends of the small and large time behaviours are very
well captured.

\subsection{Subleading terms in the Richardson regime}
\label{sub:subleading}
We have seen in previous subsections that the most noticeable effect
of inertia on the mean pair dispersion is a long transient regime that
takes place before reaching a Richardson explosive separation
(\ref{eq:r2rich}), and that this regime is due to the relaxation of
particle velocities to those of the fluid. As we now argue, at larger
times\,---\,corresponding to regime (C)\,---\,there is still an effect
of particle inertia that can be measured in terms of subleading
corrections to the Richardson law. To estimate these corrections, let
us assume that in the mean-field equation (\ref{eq:mf2}), the term
stemming from the fluid velocity $C\varepsilon^{1/3}r^{1/3}$ is much
larger than the inertia term $\tau_s \dot{v}$. This is true when
$St(r)\ll 1$, i.e.\ at times $t$ when $r(t)\gg R^*(St)$. In this
asymptotic, one can infer that the transverse velocity component $w$
is much smaller than the total velocity $v$, so that $\dot r \simeq v$
(see eq.~(\ref{eq:v-w})). In the spirit of the weak inertia expansion
derived in \cite{m87}, we next write a Taylor expansion of
(\ref{eq:mf2}) to obtain
\begin{equation}
  \dot{r} \approx v \approx C\varepsilon^{1/3} r^{1/3} - \tau_s
  \langle | (\mathrm{d}/ \mathrm{d}t)\, \delta_r \bm u
  |^2\rangle^{1/2} \approx C\varepsilon^{1/3} r^{1/3} - \tau_s \langle
  |\delta_r \bm a |^2\rangle^{1/2},
\label{eq:mf_maxeyapprox}
\end{equation}
where $\delta_r \bm a = \delta_r(\partial_t\bm u+\bm u\cdot\nabla\bm
u)$ denotes the increment of the fluid acceleration over the
separation $r$. Next we assume scaling invariance of the turbulent
acceleration field, that is, according to dimensional arguments of K41
theory, $|\delta_r \bm a| \sim \varepsilon^{2/3}r^{-1/3}$. Equation
(\ref{eq:mf_maxeyapprox}) can then be rewritten as
\begin{equation}
  \dot{r} = C\varepsilon^{1/3}r^{1/3}
  \left(1-A\,\tau_s\,\varepsilon^{1/3}r^{-2/3}\right)=
  C\varepsilon^{1/3}r^{1/3} \left(1-A\, St(r)\right),
\end{equation}
where $A$ is an order-unity constant. The initial condition is given
by $r(t_0)=r_0$ where the initial separation has to be chosen such
such that $St(r_0)\ll1$. We can next integrate the approximate
dynamics perturbatively in terms of the small parameter $St(r_0)$ by
expanding the separation as $r(t) = \rho_0(t) + \rho_1(t) + \rho_2(t)
+ \dots$. The leading order is $\rho_0(t) = [ r_0^{2/3}
+(2C/3)\,\varepsilon^{1/3} t]^{3/2}$ and corresponds to the relative
dispersion of a pair of tracers. The first-order correction reads
$\rho_1(t)= - \tau_s \,\varepsilon^{1/3} A\,\ln (\rho_0(t)/r_0) \,
\rho_0^{1/3}(t)$. At times much larger than the Batchelor time
associated to the initial separation $r_0$, i.e.\ for
$t\gg\varepsilon^{-1/3}r_0^{2/3}$, the leading term follows the
Richardson explosive law $\rho_0(t) \simeq
(2C/3)^{3/2}\varepsilon^{1/2} t^{3/2}$. This finally implies that in
the asymptotics $t \gg \varepsilon^{-1/3}r_0^{2/3} \gg \tau_s$, one
can write
\begin{equation}
  r^2(t) \propto g\, t^{3} \,\left[ 1 - D
    \left({t}/{\tau_s}\right)^{-1} \ln\left({t}/{\tau_s}\right)
    \right],
  \label{eq:correcrichard}
\end{equation}
where $g$ is the Richardson constant introduced in \S\ref{sec:2} and
$D$ is an order-unity factor, which a priori does not depend neither
on the particle Stokes number, nor on the initial particles
separation.

\begin{figure*}
\begin{center}
\includegraphics[width=0.52\textwidth]{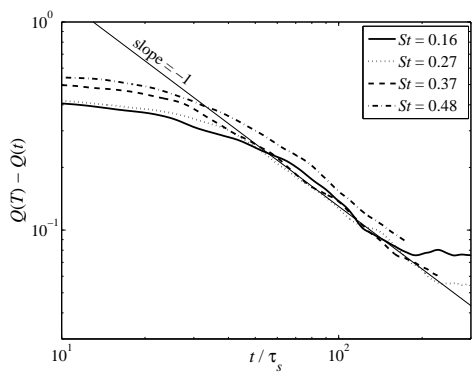}
\caption{Large-time behaviour of the mean square separation normalised
  to that of tracers as defined from Eqn. (\ref{eq:defQ}) for Run I
  and various values of the Stokes number as labeled. Deviations from
  the fluid tracer Richardson law behave as $(t/\tau_s)^{-1}$. The
  limiting value $Q(T)$ has been chosen different from unity as
  effects of inertia are still present at the largest scale of the
  flow. }
\label{fig:devrichard}
\end{center}
\vspace{-0.3truecm}
\end{figure*}

This behaviour is confirmed numerically as can be seen from
Fig.~\ref{fig:devrichard} that gives the behaviours at large times of
the ratio $Q(t)$ between the mean square separation of heavy particles
and that of tracers as defined by (\ref{eq:defQ}). One can clearly see
that data almost collapse on a line $\propto 1/t$ confirming the
behaviour (\ref{eq:correcrichard}) predicted above. Only results from
Run I are displayed here. The reason is that the very large time
statistics of tracer dispersion in Run II is not as well statistically
converged, leading to more noisy data. The qualitative picture is
however very similar.

To conclude this section, let us stress that we have assumed above K41
scaling to hold for the acceleration field (and thus for the pressure
gradient). However it is well known that the scaling properties of
pressure are still unclear: they might depend on the turbulent flow
Reynolds number and/or on the type of flow (see, e.g.,
\cite{gotoh,xu-pressure}). As stated in \cite{prl_nostro}, rather than
being dominated by K41 scaling, numerically estimated pressure
increments of Run I ($Re_{\lambda} \simeq200$) seem to be ruled by
sweeping, so that $|\delta_r \bm a|\sim
u_{\mathrm{rms}}\,\varepsilon^{1/3} r^{-2/3}$. One can easily check
that this difference in scaling leads to a behaviour similar to
(\ref{eq:correcrichard}), except that this time logarithmic
corrections are absent, and that the non-dimensional constant $D$
depends on the Reynolds number of the flow. The present numerical data
do not allow to distinguish between these two possible behaviors.

\section{Probability density function of inertial particle separation}
\label{sec:pdf}
We now discuss the shape of the probability density function for both
light and heavy inertial particles. We focus on the time and scale
behaviour of the non-stationary PDF
\begin{equation}
\label{eq:rich_st}
\cP_{St,\beta}(R,t|R_0, t_0)\,, 
\end{equation} 
defined as the probability to find a pair of inertial particles
$(St,\beta)$, with separation $R$ at time $t$, given their initial
separation $R_0$ at time $t_0$. The case of tracers ($St=0,\beta=1$)
has been widely studied in the past, either experimentally,
numerically and theoretically for two and three dimensional turbulent
flows (see \cite{rich.rev,batch.rev,tabeling,boffi.2d,bif.2p,
  bodi.science,collins.rev}).  Following the celebrated ideas of
Richardson, phenomenological modelling in terms of a diffusion
equation for the PDF of pair separation leads to the well-known
non-Gaussian distribution,
\begin{equation}
\label{eq:rich}
\cP_{St=0,\beta=1}(R,t) \propto \frac{R^2}{\left(\varepsilon^{1/3} t \right)^{9/2}}
\,\exp{\left[- \frac{A\,R^{2/3}}{\varepsilon^{1/3}\,t}\right]}\,,  
\end{equation} 
which is valid for times within the inertial range $\tau_{\eta} \ll t
\ll T_L$, and is obtained assuming a small enough initial separation
and statistical homogeneity and isotropy of the three-dimensional
turbulent flow. Here, $A$ is a normalization constant. This prediction
is based on the simple assumption that, for inertial range distances,
tracers undergo a diffusion dynamics with an {\it effective},
self-similar, turbulent diffusivity $K(R) \propto R \, \delta_R u \sim
\varepsilon^{1/3}\,R^{4/3}$. Moreover, it relies on the
phenomenological assumption that tracers separate in a short-time
correlated velocity field. Indeed, it is only if the latter is true,
that the diffusion equation for the pair separation becomes exact (see
\cite{gaw}).

As mentioned before such a scenario may be strongly contaminated by
particle inertia. The main modifications are expected to be due to the
presence of small-scale caustics for small-to-large Stokes numbers,
and to preferential concentration. Caustics make the small scale
velocity field not differentiable and not self-similar, as if inertial
particles were separating in a rough velocity field whose exponent
were depending on distance. Preferential concentration, leading to
inhomogeneous spatial distribution of particles, manifests itself as a
sort of {\it effective compressibility} in the particle velocity
field.

There exists a series of stochastic {\it toy models} for Lagrangian
motion of particles in incompressible/compressible velocity fields,
where the statistics of pair separation can be addressed
analytically. Among these, the so-called Kraichnan ensemble models,
where tracer particles move in a compressible, short-time correlated,
homogeneous and isotropic velocity field, with Gaussian spatial
correlations (we refer the reader to the review \cite{gaw} for a
description of this model). It is useful for the sequel to recall two
main results obtained for relative dispersion in a Kraichnan
compressible flow. We denote with $\wp$ the velocity field
compressibility degree\footnote{The compressibility degree $\wp$ is
  defined as the ratio $\wp\equiv {\cal C}^2/{\cal S}^2$, where ${\cal
    C}^2 \propto \langle (\nabla \cdot {\bm u})^2\rangle$ and ${\cal
    S}^2 \propto \langle (\nabla {\bm u})^2\rangle$, and varies
  between $\wp=0$ for incompressible flows, and $\wp=1$ for potential
  flows.}, and with $0 \le \xi < 2$ the scaling exponent of the
two-point velocity correlation function at the scale $r$, in
$d$-dimensions: $\la [u_i({\bm r})-u_i(0)][u_j({\bm r})-u_j(0)] \ra
\sim G_1 r^\xi [(d-1+\xi-\wp\xi) \delta_{ij}
+\xi(\wp\,d-1)r_ir_j/r^2]$. For particles moving in such flows, it is
possible to show that the pair separation PDF for tracer particles
follows a Richardson-like behaviour:
\begin{equation}
\label{eq:rich_k}
\cP_{\mu,\xi}(R,t) \propto \frac{R^{D_2-1}}{t^{(d-\mu)/(2-\xi)}}\,\exp{\left[-A
  \frac{R^{2-\xi}}{t}\right]}\,.  
\end{equation} 
Here $\mu = \wp\xi(d+\xi)/(1+\wp\xi)$ and $D_2=d-\mu$ is the correlation
dimension, characterizing the fractal spatial distribution of
particles.\\ A different distribution emerges when the $d-$dimensional
Kraichan flow is differentiable, i.e. for $\xi = 2$; in such case, a
log-normal PDF is expected:
\begin{equation}
\label{eq:rich_lognormal}
\cP_{\mu,\xi}(R,t|R_0, t_0) \propto
\frac{1}{R}\,\exp{\left[-\frac{(\log(R/R_0) - 
\lambda (t-t_0))^2}{2 \Delta (t-t_0)}\right]}\,,  
\end{equation} 
with $\Delta = 2 G_1 (d-1) (1+2\wp)$ and $\lambda =
G_1(d-1)(d-4\wp)$. It is worth noticing that in the latter case, since
the flow is differentiable, the large-time PDF depends on the initial
data.

The problem of inertial particle separation in a real turbulent flow
presents some similarities with the previous toy cases but also
important differences. \\First, the effective degree of
compressibility\,---\,due to preferential concentration of inertial
particles\,---\,, is properly defined only in the dissipative range of
scales. For $r \ll \eta$, it is equal to the correlation dimension
$D_2$ defined as $p(r) \sim r^{D_{2}}$, where $p(r)$ is the
probability to find two particles at distance smaller than $r$, with
$r\ll \eta$. As it has been numerically shown in
\cite{prl_nostro,calza} for three-dimensional turbulent flows, the
correlation dimension depends only on the degree of inertia
$(St,\beta)$, while it does not seem to depend on the Reynolds number
of the flow. For $r\gg\eta$, the effective degree of compressibility
is no longer constant, but varies with the scale.\\Second, the
underlying velocity field exhibits spatial and temporal correlations
that are much more complex than in a Gaussian short-correlated
field. Such correlations lead to non-trivial overlaps between particle
dynamics and the carrying flow topology. As a result, it is not
possible to simply translate the analytical findings obtained in the
compressible Kraichnan ensemble to the case of inertial particles: we
may expect however that in some limits the compressible Kraichnan
results should give the leading behaviour also for the case studied
here of inertial particles in real turbulent flows.

With this purpose, we first notice that the separation probability
density function that is valid in the rough case (\ref{eq:rich_k}) has
an asymptotic stretched-exponential decay that is independent on the
compressibility degree. This suggests that inertial particle PDF
(\ref{eq:rich_st}) must recover the Richardson behaviour
(\ref{eq:rich}) of tracers in the limit of large scales and large
times. Coherently with what discussed in previous sections, for large
times and for scales larger than $R^*_{St}$, we expect that the heavy
pairs (in the limit $\beta \sim 0$) PDF recovers a tracer like
distribution\,:
\begin{equation}
\cP_{St,0}(R,t) \sim \exp{\left[-A
\frac{R^{2/3}}{\varepsilon^{1/3}t}\right]}; \qquad R \gg R^*_{St}.
\label{eq:pdflight}
\end{equation}
For pairs of light particles, there is no straightforward formulation
of such a prediction: as we shall see in the sequel, preferential
concentration effects have a strong fingerprint on the separation PDF
even at large times and large scales.

In the opposite limit of very small separations, i.e.\ $R \ll \eta$,
one can correctly assume that the effective degree of compressibility
is constant and therefore apply either the small-scale limit for rough
flows (\ref{eq:rich_k}), or that for smooth flows
(\ref{eq:rich_lognormal}), depending on the scaling properties of the
particle velocity field entailed in the value of the exponent
$\gamma(St)$, defined from (\ref{eq:gamma_st}) and related to the
caustics. We thus expect
\begin{eqnarray}
  \cP_{St,\beta}(R,t|R_0, t_0) &\sim& R^{\frac{D_2}{2} -1} G(t),
  \qquad \mbox{if }\gamma(St)\neq 1, \nonumber
  \\ \cP_{St,\beta}(R,t|R_0, t_0) &\sim& R^{D_2-1} F(t), \qquad
  \mbox{if }\gamma(St)= 1.
\label{eq:log}
\end{eqnarray} 
Here $F$ and $G$ are two different decaying functions of time $t$,
whose expression can be easily derived from
(\ref{eq:rich_k})-(\ref{eq:rich_lognormal}). Notice that for the
smooth case, i.e. the small-scale limit of the log-normal distribution
(\ref{eq:rich_lognormal}), we get for the spatial dependency a factor
$D_2/2$ instead of the factor $D_2$ of the rough case. This will
matter in the case of light particle separation, where, due to strong
preferential concentration, the probability of finding pairs at a very
small distances is large enough to allow for a detailed test of the
prediction (\ref{eq:log}). The case of light particles will be
discussed in \S\ref{sec:light}, while we now turn to a discussion of
the above scenario in the case of heavy particle pairs.

\subsection{Probability density function of heavy particle relative
  separation}
\label{subsec:pdfheavy}
\begin{figure*}
\begin{center}
\includegraphics[width=1.\textwidth]{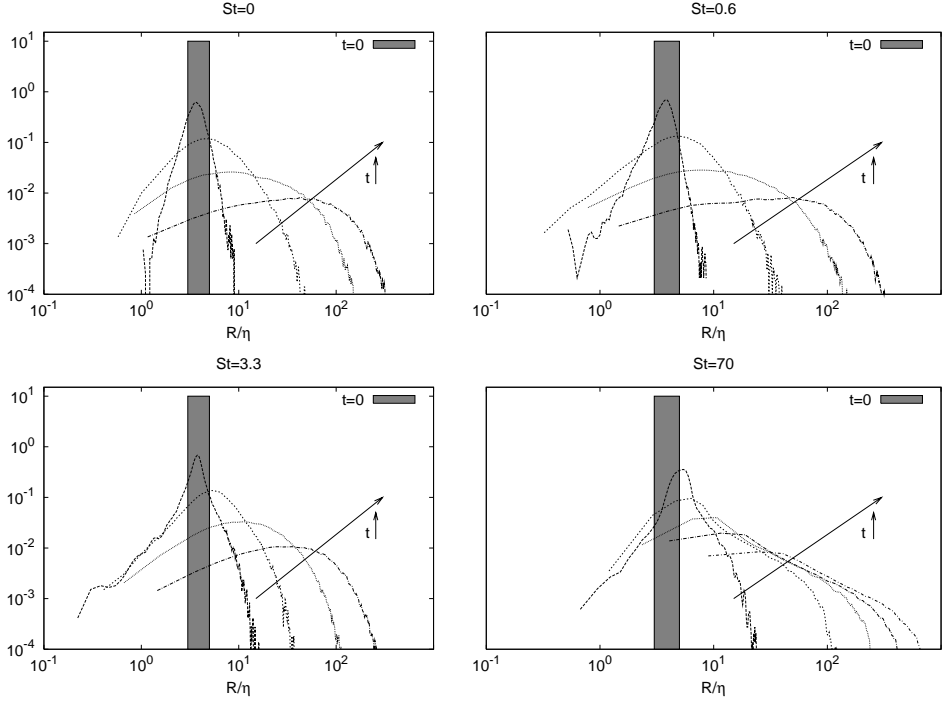}
\caption{Separation probability density function, $\cP(R,t|R_0,t_0)$,
  for heavy pairs with different Stokes numbers, at changing
  time. Initial distance is taken $R_0 \in [3\!:\!4]\,\eta$ for $St=0$
  (top-left), $St = 0.6$ (top-right), and $St = 3.3$ (bottom-left)
  of Run I, and equal to $R_0 \in [4\!:\!6]\,\eta$ for $St=70$
  (bottom-right) of Run II. The related initial distributions are
  pictorically depicted with a grey area. Times shown are:
  $(t-t_0)/\tau_\eta=1,6,18,36$ for Run I and
  $(t-t_0)/\tau_\eta=1,6,18,36,86$ for Run II.}
\label{fig:1}
\end{center}
\vspace{-0.3truecm}
\end{figure*}
\begin{figure*}
\begin{center}
\includegraphics[width=1.\textwidth]{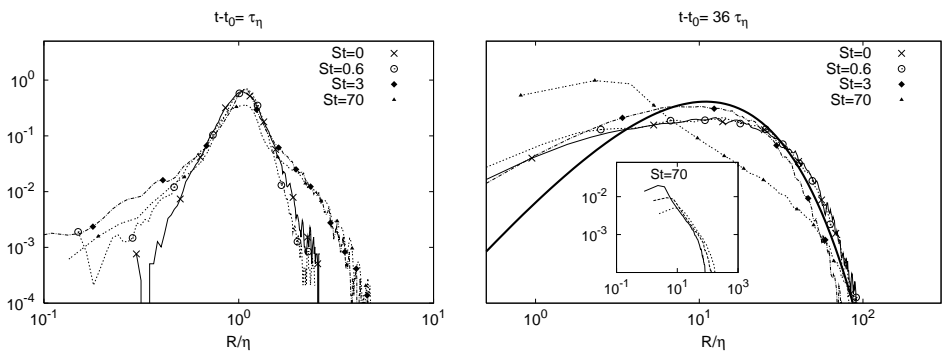}
\caption{Comparison of PDFs at fixed times with data of
  Fig.(\ref{fig:1}). Left: early stage of the separation process,
  $t-t_0 = \tau_\eta$. Inertia does not affect small Stokes, $St=0.6$
  while its effect is detectable for $St=3$ and $St=70$. Right: PDFs
  comparison at a later time, $t-t_0 = 36 \tau_\eta$. Now the PDF
  shows some deviations from the tracer behaviour only for $St=70$. On
  the right panel the solid line gives the Richardson shape
  (\ref{eq:rich}). Initial separation and Reynolds numbers are the
  same as for Fig. (\ref{fig:1}). The inset shows the PDF evolution
  for $St=70$ at three times, $(t-t_0)/\tau_\eta= 36,82,130$.}
\label{fig:2}
\end{center}
\vspace{-0.3truecm}
\end{figure*}

We start by analyzing the qualitative evolution of $\cP_{St}(R,t)$ at
changing time, for different Stokes numbers and in the limit
$\beta=0$. The four panels of Fig.~\ref{fig:1} show the evolution of
the PDF at different times for pairs with $St=0$ (tracers), and for
heavy particles with $St=1$, $3.3$, and $70$. Initially, at $t=t_0$,
all selected pairs are separated by the same distance ($R_0 \in
[3\!:\!6]\,\eta$); this initial distribution is represented in each
figure by a grey area. As time elapses, particle separate and reach
different scales, depending on their inertia. Qualitatively, the PDF
evolution is very similar for all moderate Stokes numbers, and the
PDFs at different moderate Stokes numbers become more and more similar
with time. However in the case of $St=70$\,---\,for which the
associate Stokes time $\tau_s$ falls well inside the inertial
range\,---\,, the PDF shows a long exponential tail for intermediate
separation, which tends to persist at all observed times. To better
appreciate such differences, in Fig.\ \ref{fig:2} we show the
comparison between the different PDFs corresponding to various Stokes
numbers for two different times: at the beginning of the separation
process, $(t-t_0) = \tau_\eta$, and at a later time, $(t-t_0)=36
\tau_\eta$.
As one can see, it is only at early times that the PDFs for
moderate-to-large Stokes, $St=3,70$ differ in a sensible way from the
tracers. In particular, one can clearly see that many pairs have
separations much larger and much smaller than those observed for
tracers or for heavy pairs with small Stokes numbers. The right tails,
describing pairs that are very far apart, are just the signature of
the {\it scrambling} effect of caustics. Such strong events are not
captured by second order moments of the separation statistics that we
discussed before, while they clearly affect higher-order moments.  The
left tails, associated to pairs much closer than tracers, are possibly
due to particles that separate at a slower rate than tracers because
of preferential concentration induced by inertia.

Later in the evolution, for $(t-t_0)=36 \tau_\eta$, only the
separation PDF for $St=70$ still shows important departure from the
tracer case; for all the other Stokes numbers shown, pairs have had
enough time to forget their initial distribution and have practically
relaxed on the typical Richardson-like distribution. In the inset, we
also show the persistence in the exponential behaviour for the PDF at
$St=70$, by superposing the shapes measured at three times during the
particles separation.

With the present data the small scale asymptotic behaviour
(\ref{eq:log}) cannot be validated for heavy particles. This is due to
the limited statistics: very soon after the initial time $t_0$, there
are almost no pairs left with separations $R \ll \eta$.

\subsection{Probability density function of heavy particle relative
  velocities}
\label{subsec:caustics}
At moderate to large Stokes numbers the separation process of heavy
particle pairs is largely influenced by the presence of large velocity
differences at small scales, that is by the presence of caustics in
the particles velocity field. In \S\ref{sec:1}, we have studied
stationary statistics (only first-order moment) of velocity
differences between heavy particles at changing the distance between
particles and their inertia. However it is also informative to look at
the non-stationary, time-dependent distribution of velocity
differences, and more particularly to its distribution measured along
heavy pairs separation. The relative velocity $ \delta_R{\bm V}(t) =
\dot{\bm X}_1(t) -\dot{\bm X}_2(t)$ can be decomposed into the
projection along the separation vector, and two transveral components,
here equivalent since the system is statistically isotropic. For
tracer particles, the statistics of relative velocity and the
alignment properties of $ \delta_R{\bm V}(t)$ and $\bm R(t)$ have been
discussed extensively (see e.g.\ \cite{YB}). Here, we focus on the PDF
of the relative longitudinal velocity only, which we denote by
$\cW_{St}(v,t)$, where $v(t) = {[ \dot{\bm X}_1(t) -\dot{\bm X}_2(t)]
  \cdot \hat{\bm R}(t)}$.
\begin{figure*}
\begin{center}
\includegraphics[width=1.\textwidth]{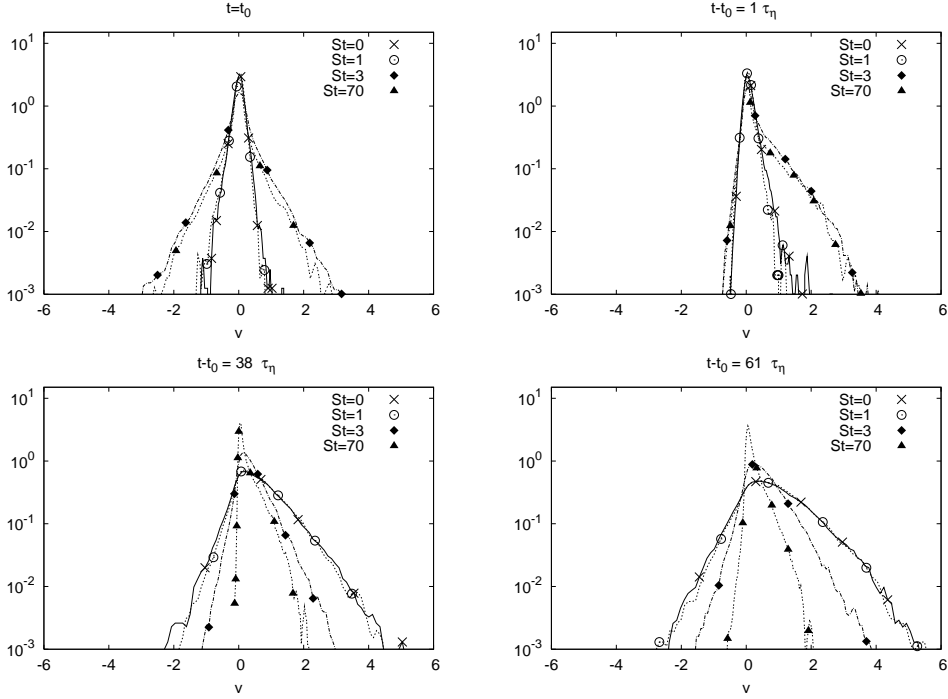}
\caption{Time evolution of the probability density function of heavy
  particle relative longitudinal velocity, $\cW_{St}(v,t)$, during
  the separation process. Data refers to four different cases: tracers
  pairs $St=0$, and heavy pairs $St=1,3,70$, starting with initial
  distance $R_0 \in [4:6] \eta$. PDFs are measured at times
  $(t-t_0)=[0,1,38,61] \tau_{\eta}$, for Run II. Notice the presence
  of intense velocity fluctuations for moderate-to-strong inertia,
  $St=3, 70$, observable at the early stage of the separation
  process. These are the legacy of the caustics distribution.}
\label{fig:4}
\end{center}
\vspace{-0.3truecm}
\end{figure*}
For pairs of tracers ($St=0$), the initial longitudinal velocity
distribution is nothing else than the PDF of Eulerian longitudinal
velocity increments measured at the distance $R_0$.  For pairs of
inertial particles, this initial PDF clearly coincides with the
stationary distribution of velocity differences between particles that
are at a distance $R=|\bm X_1(t_0)-\bm X_2(t_0)|\in
[R_0\!:\!R_0+dR_0]$. Such a distribution has the signature of two
mechanisms: (\textit{i}) at small Stokes numbers, only preferential
concentration matters and particles probe only a sub-set of all
possible fluid velocity fluctuations; (\textit{ii}) at large Stokes
numbers, particles are homogeneously distributed but with a velocity
field which may be strongly different from the underlying fluid
velocity.

For what concerns heavy pairs, the first effect has not an important
signature on small-scale quantities. However the second effect clearly
becomes visible for moderate to high inertia as shown in
Fig.~\ref{fig:4}. Here we report the longitudinal velocity
distributions for pairs with initial distance $R_0 \in [4\!:\!6]\,
\eta$, and with $St=0,1,3.3,$ and $70$; the Reynolds number of the
underlying flow is $Re_{\lambda} \sim 400$. Each panel contains the
PDFs measured at different times spanning all turbulent timescales.
At $t=t_0$, the importance of caustics is manifest for the two largest
Stokes numbers, leading to fat tails towards both small and large
velocity differences. Interestingly enough, the left tail of
$\cW_{St}(v,t)$, which describes approaching events of particle
relative motion, is immediately dumped already at $ (t-t_0) \sim
\tau_\eta$; at the same time, however, the right tail continues to be
quite fat for the two largest Stokes numbers under consideration. At
later stages of the separation process, the tendency of
large-Stokes-number pairs to wash out approaching events becomes even
stronger. Indeed, at time $(t-t_0)=38\,\tau_{\eta}$, the small
velocity increments tail has almost disappeared for pairs with
$St=70$. It is worth noticing that at those times (i.e.\ also at those
typical scales), heavy particle velocity differences have already
started to be smaller than the tracer velocity increments: the larger
is the Stokes number, the less pronounced are the PDF tails.

Summarising, because of the different effects of inertia, we observe a
very complex evolution for the longitudinal relative velocity
fluctuations along the trajectories of heavy particle pairs. This is
certainly a key issue to be considered for stochastic modelling: here,
as in a standard {\it kinetic} problem, both particle positions and
velocities need to be modelled to quantitatively control the relative
dispersion process.

\section{Relative dispersion for light particles}
\label{sec:light}
So far we have considered the relative motion of very heavy particle
pairs, for which the density contrast $\beta$ with the underlying
fluid is zero. In this section we present results on light particles
dynamics as described by (\ref{eq:1}), for different possible choices
of the parameters $(St,\beta)$.

\begin{figure*}
\begin{center}
\includegraphics[width=0.52\textwidth]{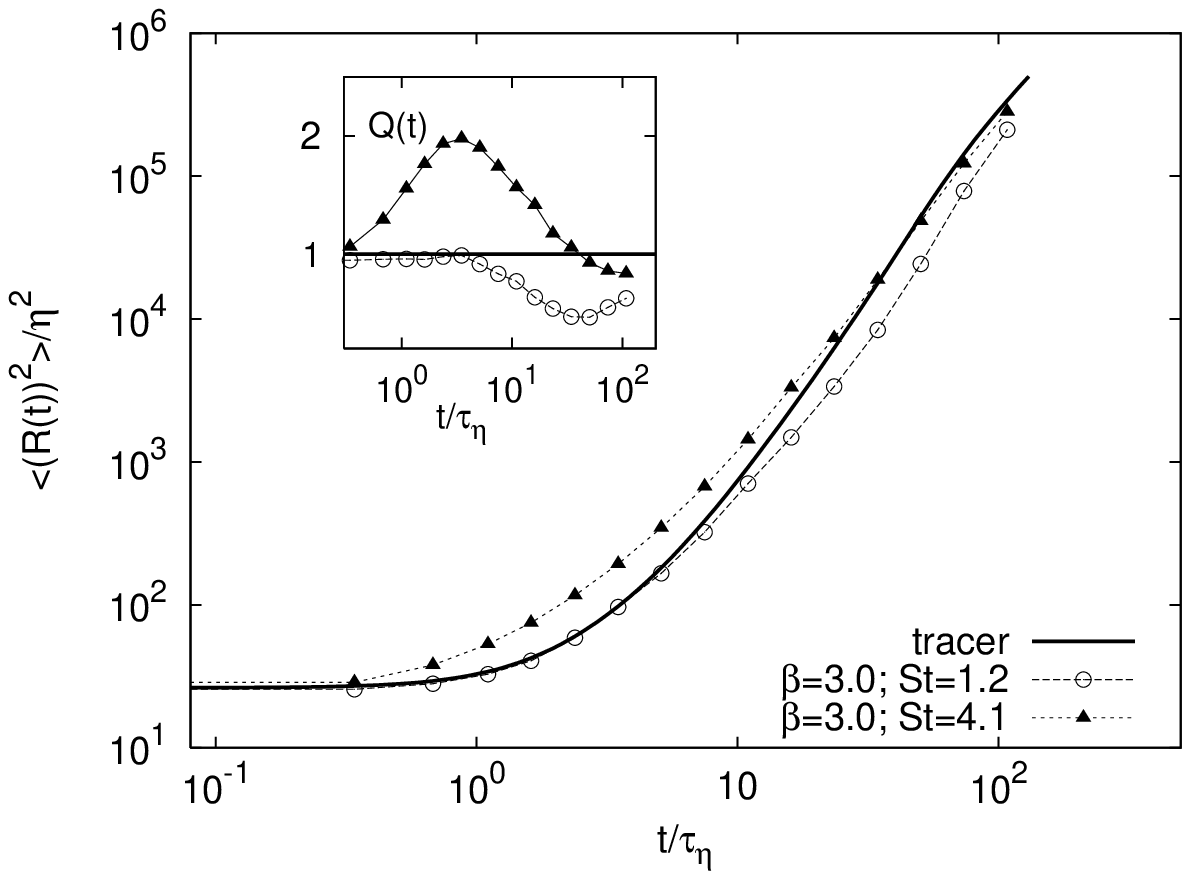}
\vspace{-10pt}
\caption{Time evolution of the mean square separation for two
  different families of light particles $(St=1.2,\beta=3)$ and
  $(St=4.1,\beta=3)$. The case of tracers is also shown for
  comparison. Notice that the strong small-scale clustering does
  not affect the long-time behaviour, except through a very small
  asymptotic slow down. Inset: ratio between the mean square
  separation for light pairs and that of tracers.}
\label{fig:rich.light}
\end{center}
\vspace{-0.3truecm}
\end{figure*}
We discuss how the strong effect of preferential
concentration\,---\,typically observed in the case of light particles
in turbulent flows\,---\, might influence the intermediate and
long-time behaviour of pair separations. In three-dimensional
turbulent flows, as we consider here, light particles associated to
different values of $(St,\beta)$ have been observed to always possess
a positive largest Lyapunov exponent~\cite{calza}: this implies that
light pairs always separate in $3d$ real turbulent flows. We recall,
however, that this is not always true: for instance, in smooth
two-dimensional random flows, there are values of $(St,\beta)$ for
which the largest Lyapunov exponent can become negative and particles
form pointwise clusters (see \cite{bec2003}).

Light particles with moderate inertia and high density ratio
(order-unity $St$, and $\beta =3$), initially tend to separate much
slower than heavy particles with the similar Stokes number: this is
evident from the much smaller values of the Lyapunov exponents
measured for light particles, with respect to those measured for heavy
particles with equivalent Stokes numbers but $\beta=0$. Moreover,
finite-time Lyapunov exponents show large fluctuations, indicating
that there are pairs that do not separate even at long times. Results
on this issue will be reported elsewhere. Clearly, pairs of light
particles that do not separate do not influence the mean square
distances: hence, we do not expect, and indeed do not measure, any
large differences for the long-time behaviour of $\langle |\bm R(t)|^2
\rangle_{St,\beta}$ for light particles, with respect to the heavy
case (see Fig.\ \ref{fig:rich.light}).

\begin{figure*}
\begin{center}
\includegraphics[width=1.\textwidth]{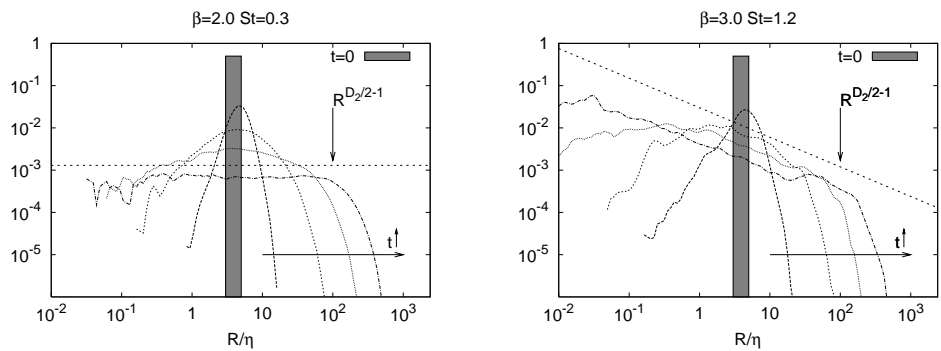}
\vspace{-5pt}
\caption{Time evolution of the separation probability density
  functions of light particles. Left panel: $St=0.3, \beta=2$,
  corresponding to a case where preferential concentration is not very
  effective ($D_2 = 2$).  As time elapses, one observes a self-similar
  filling toward smaller separations, in agreement with (\ref{eq:log})
  (dashed line). Right panel: $St=1.2,\beta=3$, corresponding to light
  particles with strong clustering properties (correlation dimension
  $D_2=0.8$).  Again the self-similar filling of small scales is
  consistent with the prediction $D_2/2-1$ as depicted by the dashed
  straight line.}
\label{fig:light1}
\end{center}
\vspace{-0.3truecm}
\end{figure*}

It is natural to ask if light particle strong preferential
concentration affects high-order moments of relative separation of two
initially close particles, and particularly the left tail of the
separation PDF. Figure~\ref{fig:light1} shows the time evolution of
the separation probability density function, $\cP_{St,\beta}(R,t|R_0,
t_0)$. Data refer to a case with very intense preferential
concentration effects and minor influence of caustics
$(St=1.2,\beta=3)$, and a case with milder inhomogeneities in the
spatial distribution $(St=0.3,\beta=2)$. The initial separation PDF
was chosen in both cases by selecting particle pairs with initial
distance $R_0 \in [4\!:\!6]\,\eta$. A remarkable observation is a
strong tendency to fill small separations. In other words there are
many pairs that reduce their mutual distance even for a very long
times. The development of the left tail for the strong clustering case
($St=1.2,\beta=3$) is consistent with the estimate given by the
long-time, small-scale asymptotic expansion of the log-normal
distribution (\ref{eq:log}), $\cP(R,t) \sim R^{D_2/2-1}$, as shown by
the straight line in the plot. This is the confirmation that the
small-scale dynamics of the highly clustered light particles evolves
as that of tracers moving in a smooth, compressible flow
(characterised by the same $D_2$).

We also remark that if there is high spatial preferential
concentration, caustics cannot be important. This may have important
consequences for the estimation of collision kernel of light
particles. The approaching events, shown by the left tail in
Figure~\ref{fig:light1}, are clearly due to the preferential
concentration inside vortex-like structures, typical of light
particles. The right panel show a different case, where preferential
concentration is less important, leading to a correlation dimension
$D_2=2$. Of course, also in this latter case, there are events with
approaching pairs, but these become less and less probable with time.

\begin{figure*}
\begin{center}
\includegraphics[width=1.\textwidth]{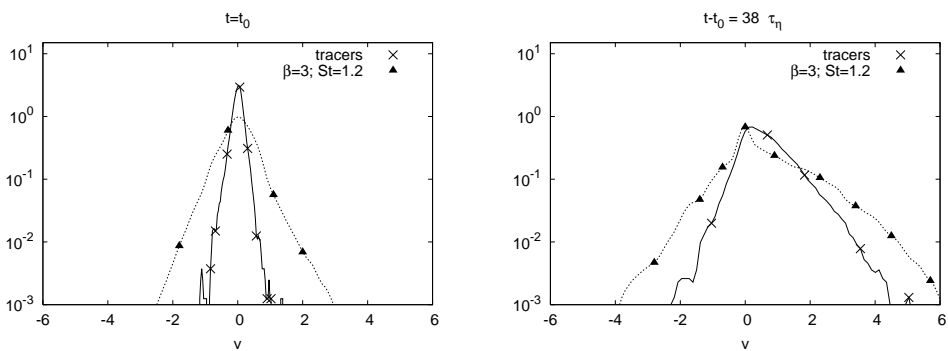}
\caption{Probability density functions of relative longitudinal
  velocity, $\cW_{St,\beta}(v,t)$, for light pairs with
  $St=1.2,\beta=3$, and for tracers ($St=0,\beta=1$). PDFs are
  measured along the separation process at two different times: left
  panel refers to the initial time, $t=t_0$; while right panel refers
  to $t-t_0=38 \tau_{\eta}$. }
\label{fig:light2}
\end{center}
\vspace{-0.3truecm}
\end{figure*}

The importance of preferential concentration can also be appreciated
by looking at the PDFs of longitudinal velocity differences between
light particles during the separation. We show such distributions for
one of the pair family considered above, and we compare them with
those of the tracers (see Fig.~\ref{fig:light2}). The important
difference between the two cases stems from the highly peaked nature
of the relative velocity PDF for the strong clustered light
particle case. The presence of many pairs with almost vanishing
velocity differences is the signature of a coherent bunch of pairs
moving on a strongly clustered set.

\section{Conclusions}
\label{sec:conc}
We have studied the relative dispersion of inertial particles in
homogeneous and isotropic turbulence from two DNS at resolutions
$512^3$ and $2048^3$, corresponding to $Re_\lambda \sim 200 $ and
$Re_\lambda \sim 400$, respectively.  We have analysed both heavy and
light particle statistics at changing the Stokes numbers. We have
studied the evolution of mean separations and the whole PDFs' shape,
both for particle distance and velocity increments at changing time
and for different typical initial distances. The main results that we
have discussed can be summarised as follows. Separations of very heavy
particles, with Stokes times falling in the inertial range of the
underlying fluid, are strongly affected by the presence of caustics up
to times, when the distance between particles reaches scales that are
large enough for the separation dynamics to be again dominated by the
underlying flow velocity. As a consequence, strong transient departure
from the Richardson diffusion, with a faster ballistic regime, is
observed.  A statistical closure of the equation of motions for heavy
particle separation is also developed. This model is able to reproduce
the main numerical findings. \\For light particles, at high density
ratio, we observe strong small-scale clustering properties, leading to
a considerable fraction of pairs that do not separate at
all\,---\,although the maximum Lyapunov exponent remains positive. In
such a case, the non-stationary spatial concentration at small scales
tends to be higher than the analogous case but with a stationary
distribution of particles. Such numerical findings open the way to
experimental verifications and gives input to the community involved
in modelling inertial particle diffusion in applied configurations.

\begin{acknowledgments} 
  We thank Massimo Cencini, who collaborated at an early stage of this
  work, for very fruitful and continuous exchange of ideas. This study
  benefitted from constructive discussions with E.\ Bodenschatz, E.\
  Calzavarini, L.\ Collins and G.\ Falkovich to whom we wish to
  express a warm gratitude. J.\ Bec and A.S.\ Lanotte aknowledge
  support from the National Science Foundation under grant No.\
  PHY05\_51164 for their stay at the Kavli Institute for Theoretical
  Physics in the framework of the 2008 ``Physics of the Climate
  Change'' program. Part of this work was supported by Agence
  Nationale de la Recherche under grant No.\ BLAN07-1\_192604. We
  thank the DEISA Consortium (co-funded by the EU, FP6 project
  508830), for support for Run II within the DEISA Extreme Computing
  Initiative (www.deisa.org).  Numerical simulations of Run I were
  performed at CASPUR (Italy) and CINECA (Italy).  Numerical raw data
  of particle trajectories are freely available from the iCFDdatabase
  (http://cfd.cineca.it) kindly hosted by CINECA (Italy).
\end{acknowledgments}

\end{document}